%% file: mst_methods_paper.tex
\documentclass[fleqn,usenatbib]{mnras}

\usepackage{newtxtext,newtxmath}

\usepackage[T1]{fontenc}
\usepackage{ae,aecompl}

\usepackage{graphicx}	
\usepackage{amsmath}	
\usepackage{amssymb}	
\usepackage{amsbsy}
\usepackage{siunitx}
\usepackage{tabularx}
\usepackage{breqn}
\usepackage{xcolor}

\DeclareSIUnit \h {h}
\DeclareSIUnit \parsec {pc} 
\DeclareSIUnit \ra {RA}
\DeclareSIUnit \dec {Dec}
\DeclareSIUnit \redshift {z}
\DeclareSIUnit \solarmass {M_{\odot}}
\DeclareSIUnit \sig {\sigma}
\def\mathbi#1{\textbf{\em #1}}

\newcommand\Tstrut{\rule{0pt}{2.9ex}}       
\newcommand\Bstrut{\rule[-1.3ex]{0pt}{0pt}} 
\newcommand\TBstrut{\Tstrut\Bstrut}         

\title[Beyond 2PT: using the MST for cosmology]{Beyond two-point statistics: using the minimum spanning tree as a tool for cosmology}

\author[K. Naidoo et al.]{Krishna Naidoo$^{1}$\thanks{E-mail: \href{mailto:krishna.naidoo.11@ucl.ac.uk}{krishna.naidoo.11@ucl.ac.uk}},
	Lorne Whiteway$^{1}$, Elena Massara$^{2}$, Davide Gualdi$^{3}$, Ofer Lahav$^{1}$,
    \newauthor{Matteo Viel$^{4,5,6,7}$, H\'{e}ctor Gil-Mar\'{i}n$^{3,8}$ and Andreu Font-Ribera$^{1}$}\\
	\\
	$^{1}$Department of Physics and Astronomy, University College London, Gower Street, London WC1E 6BT, UK\\
	$^{2}$Center for Computational Astrophysics, Flatiron Institute, 162 5th Avenue, New York, NY 10010 USA\\
	$^{3}$ICC, University of Barcelona, IEEC-UB, Mart\'{i} i Franqu\`{e}s, 1, E-08028 Barcelona, Spain\\
	$^{4}$SISSA -- International School for Advanced Studies, Via Bonomea 265, I-34136 Trieste, Italy\\
	$^{5}$INAF -- Osservatorio Astronomico di Trieste, Via G. B. Tiepolo 11, I-34143  Trieste, Italy\\
	$^{6}$INFN -- National Institute for Nuclear Physics, via Valerio 2, I-34127 Trieste, Italy\\
	$^{7}$IFPU -- Institute for Fundamental Physics of the Universe, via Beirut 2, I-34151 Trieste, Italy\\
	$^{8}$Institute of Space Studies of Catalonia (IEEC), E-08034 Barcelona, Spain\\
}

\date{Accepted XXX. Received YYY; in original form ZZZ}

\pubyear{2019}

\begin{document}
\label{firstpage}
\pagerange{\pageref{firstpage}--\pageref{lastpage}}
\maketitle

\begin{abstract}
Cosmological studies of large-scale structure have relied on two-point statistics, not fully exploiting the rich structure of the cosmic web. In this paper we show how to capture some of this cosmic web information by using the minimum spanning tree (MST), for the first time using it to estimate cosmological parameters in simulations. Discrete tracers of dark matter such as galaxies, $N$-body particles or haloes are used as nodes to construct a unique graph, the MST, that traces skeletal structure. We study the dependence of the MST on cosmological parameters using haloes from a suite of COLA simulations with a box size of $250\ h^{-1}{\rm Mpc}$, varying the amplitude of scalar fluctuations $\left(A_{\rm s}\right)$, matter density $\left(\Omega_{\rm m}\right)$, and neutrino mass $\left(\sum m_{\nu}\right)$. The power spectrum $P$ and bispectrum $B$ are measured for wavenumbers between $0.125$ and $0.5$ $h{\rm Mpc}^{-1}$, while a corresponding lower cut of $\sim12.6$ $h^{-1}{\rm Mpc}$ is applied to the MST. The constraints from the individual methods are fairly similar but when combined we see improved $1\sigma$ constraints of $\sim 17\%$ ($\sim 12\%$) on $\Omega_{\rm m}$ and $\sim 12\%$ ($\sim 10\%$) on $A_{\rm s}$ with respect to $P$ ($P+B$) thus showing the MST is providing additional information. The MST can be applied to current and future spectroscopic surveys (BOSS, DESI, Euclid, PSF, WFIRST, and 4MOST) in 3D and photometric surveys (DES and LSST) in tomographic shells to constrain parameters and/or test systematics.
\end{abstract}





\input{introduction} 

\input{method} 

\input{sensitivity} 

\input{systematics} 

\input{nbody} 

\input{comparison} 

\input{discussion} 

\section*{ACKNOWLEDGEMENTS}

We thank Donnacha Kirk for his contributions to the early stages of this project and Niall Jeffrey for useful discussions.

Many of the figures in this paper were made using {\sc matplotlib}\footnote{\href{https://matplotlib.org/index.html}{https://matplotlib.org/index.html}} \citep{matplotlib}, whilst the corner plots were made using {\sc ChainConsumer}\footnote{\href{https://samreay.github.io/ChainConsumer/index.html}{https://samreay.github.io/ChainConsumer/index.html}} \citep{chainconsumer}.

KN acknowledges support from the Science and Technology Facilities Council grant ST/N50449X. DG acknowledges support from European Union's Horizon 2020 research and innovation programme ERC (BePreSySe, grant agreement 725327), Spanish MINECO under projects AYA2014-58747-P AEI/FEDER, UE, and MDM-2014-0369 of ICCUB (Unidad de Excelencia Mar\'{i}a de Maeztu). OL acknowledges support from a European Research Council Advanced Grant FP7/291329 and from an STFC Consolidated Grant ST/R000476/1. MV is supported by INFN PD51 INDARK grant. AFR was supported by an STFC Ernest Rutherford Fellowship, grant reference ST/N003853/1.




\bibliographystyle{mnras}
\bibliography{bib} 




\appendix

\input{appendix} 

\end{document}

%% file: introduction.tex
\section{Introduction}

Over the years, a series of probes have emerged as standard tools for cosmological parameter inference. Surveys of the cosmic microwave background (CMB), large-scale structure (LSS), weak lensing (WL), and distance ladder have dominated our knowledge of cosmological parameters through measurements of the CMB angular power spectra \citep[e.g.][]{PlanckPara2018}, galaxy clustering \citep[e.g.][]{Arthur2019}, weak lensing \citep[e.g.][]{DESWeakLensing2017, KIDS2017}, Baryonic Acoustic Oscillations (BAO) from galaxies \citep[e.g.][]{BOSSBAO2016} and Lyman alpha \citep[e.g.][]{eboss2019}, standard candles \citep[e.g.][]{Riess2016} and, more recently, standard sirens \citep[e.g.][]{StandardSirens2017}. These techniques are relatively mature, well understood and most importantly, reliable and trusted.

However, many of these techniques (but not all) rely on measuring the two-point correlation function (2PCF) or its Fourier space equivalent, the power spectrum. 
Studies that include higher order statistics, such as the three-point correlation function \citep[e.g.][]{threepoint2017} or bispectrum \citep[e.g.][]{Hector2017}, have already provided interesting constraints on cosmological parameters, demonstrating the need to go beyond the 2PCF. Despite solutions to improve the speed of 2PCF and 3PCF estimators \citep[see][]{Scoccimarro2015, Zachary2016}, going beyond the 3PCF is currently computationally intractable. The computational cost of current N-point correlation functions (NPCF) estimators scales by $\mathcal{O}(n^{N})$; for this reason this information remains to be exploited.

The most attractive reason to explore methods that incorporate higher order statistics is their potential to break existing parameter degeneracies, to provide tighter constraints and to test systematics. Of growing interest to cosmologists is the total mass of the three neutrino species, $\sum m_{\nu}$. Neutrinos are massless in the standard model of particle physics; however this cannot be the case since neutrinos oscillate \citep{Kamiokande1998, Sudbury2001}. Fortunately, LSS is sensitive to the mass of these elusive particles. As neutrinos are very light they possess high thermal velocities and dampen structure formation at scales below the free streaming scale (set by when they become non-relativistic). This effect is dependent on $\sum m_{\nu}$ and although it can be measured, the effect is small and highly degenerate with the matter density ($\Omega_{\rm m}$) and the variance of density perturbations (e.g. as measured at $8h^{-1}{\rm Mpc}$ ($\sigma_{8}$)). Currently, upper bounds of $\sum m_{\nu} \lesssim 0.12-0.23$ \si{\electronvolt} (95\% confidence limits) \citep{BOSSLYA2015, PlanckPara2016, BOSSBAO2016, Arthur2019} have been established from cosmology (specifically CMB and galaxy surveys) whilst the lower bound of $\gtrsim0.06$ \si{\electronvolt} is given by neutrino oscillation experiments. Future experiments will be able to go further; in particular experiments such as the \emph{Dark Energy Spectroscopic Instrument} \citep[DESI][]{DESI2016} are expected to probe below the lower bound of $\sim 0.06$ \si{\electronvolt}, and are expected to make a detection of the neutrino mass \cite[see][]{Font2014}. However, this is to be achieved purely by a more precise measurement of the 2PCF, not by the inclusion of extra information. 

We know from $N$-body simulations that the universe at late times appears as a cosmic web \citep{Bond1996}. Currently this cosmic web structure is not fully incorporated into the inference of cosmological parameters. In this paper we turn to graph theory, looking specifically at the minimum spanning tree (MST), to try to capture some of this rich information.
The MST was first introduced to astronomy by \citet{Barrow1985}. It has been typically used in cosmology for LSS classification, for example to search for cosmic web features such as filaments \citep[see][]{Bhavsar1988, Pearson1995,Krzewina1996,Ueda1997,Coles1998,Adami1999,Doroshkevich1999,Doroshkevich2001,Colberg2007,Balazs2008,Park2009,Adami2010,Demianski2011,Durret2011,Cybulski2014,GAMA2014,Shim1,Shim2,Shim3,Beuret2017,Campana2018,Campana2018b,Libeskind2018,Clarke2019}. It has also been used in other contexts such as determining mass segregation in star clusters \citep{Allison2009} and the generalized dimensionality of data points, fractals and percolation analysis \citep[see][]{Martinez1990,vandeweygaert1992,Bhavsar1996}. More recently, the MST was used in particle physics to distinguish between different classes of events in collider experiments \citep{Rainbolt2016}. The MST's strength is in its ability to extract patterns; this is precisely why it has been used to extract cosmic web features (the type of information currently missing from most cosmological studies). The MST's weaknesses are that the statistics cannot be described analytically and that they depend heavily on the density of the tracer. This means any comparison of models via the MST will be dependent on simulations. While this makes parameter inference more challenging, the reliance on simulations is not new; in fact parameter inference through artificial intelligence (AI) and machine learning (ML) will be similarly reliant. Here, the MST may provide a bridge between the traditional 2PCF and AI/ML, allowing us to understand the information being extracted by these AI/ML algorithms.

Our goal in this paper is to understand whether the MST could be a useful tool for cosmological parameter inference for current or future photometric and spectroscopic galaxy redshift surveys. These include the \emph{Baryon Oscillation Spectroscopic Survey},\footnote{\href{http://www.sdss3.org/surveys/boss.php}{http://www.sdss3.org/surveys/boss.php}} \emph{Dark Energy Survey},\footnote{\href{http://www.darkenergysurvey.org}{http://www.darkenergysurvey.org}} DESI,\footnote{\href{http://desi.lbl.gov/}{http://desi.lbl.gov/}} \emph{Large Synoptic Survey Telescope},\footnote{\href{https://www.lsst.org/}{https://www.lsst.org/}} \emph{Euclid},\footnote{\href{http://www.euclid-ec.org/}{http://www.euclid-ec.org/}} \emph{Prime Focus Spectrograph},\footnote{\href{https://pfs.ipmu.jp/index.html}{https://pfs.ipmu.jp/index.html}} \emph{Wide Field Infrared Survey Telescope},\footnote{\href{https://wfirst.gsfc.nasa.gov/}{https://wfirst.gsfc.nasa.gov/}} and \emph{4-metre Multi-Object Spectroscopic Telescope}.\footnote{\href{https://www.4most.eu/cms/}{https://www.4most.eu/cms/}} With this in mind, the paper is organized as follows. In Section \ref{method}, we describe the MST construction and statistics and we summarize the suites of simulations used in later Sections. In Section \ref{higherorder}, we demonstrate that the MST is sensitive to higher order statistics (i.e. beyond two-point). In Section \ref{systematics}, we explore relevant sources of systematics and methods to mitigate them. In addition, we test the sensitivity to redshift space distortions (RSDs). In Section \ref{section_nbody}, we explore the MST statistics on an unbiased tracer, and try to determine what the MST is actually measuring about the underlining density distribution. Lastly, in Section \ref{comparison}, we compare the MST's constraining power to that of the more traditional power spectrum and bispectrum measurements.

\begin{figure*}
	\includegraphics[width=\textwidth]{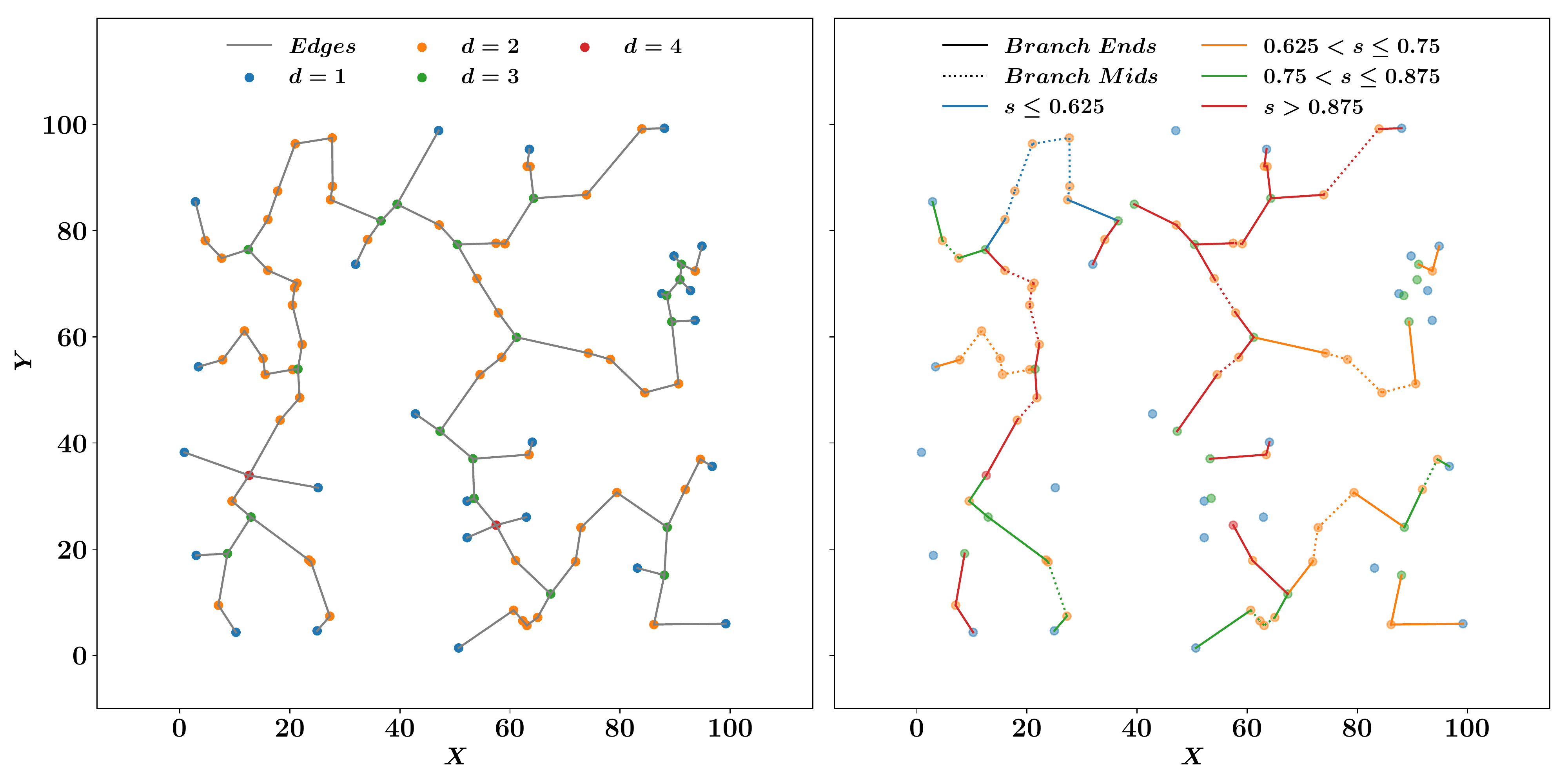}
	\caption{The MST constructed from 100 random points. Left-hand panel: the MST edges are shown. Nodes are colour coded according to their degree, i.e. the number of edges attached to them. Right-hand panel: the MST branches are colour coded according to their branch shape parameter (s). Edges that form branch ends are indicated by solid lines while edges forming the middle of branches (branch mids) are indicated by dotted lines.}
	\label{fig:mst_example}
\end{figure*}

%% file: method.tex
\section{Method}
\label{method}

In this section, we will describe: 
\begin{enumerate}
	\item Some basic properties of graphs and the MST.
	\item How the MST is constructed.
	\item The statistics we measure.
	\item Techniques for error estimation.
	\item The simulations used in this paper.
\end{enumerate}

In mathematics, a \textit{graph} is a set of \textit{nodes} (points) together with a set of \textit{edges}, where each edge joins two distinct nodes; given any two distinct nodes, there will be either zero or one edge between them. In this paper, all graphs are \textit{undirected} and \textit{weighted} i.e. an edge does not have an orientation, but it does have a (positive) weight (which in this paper will be the distance (defined below) between the nodes that it connects). A \textit{path} is a sequence of nodes in which each consecutive pair of nodes is connected by an edge (and no edge is used twice); a path that returns to its starting point is a \textit{cycle}. If there is an edge (respectively path) between any two distinct nodes then the graph is \textit{complete} (respectively \textit{connected}). Given a connected graph $G$ (not necessarily complete), one can discard edges to obtain the MST of $G$. By definition this new graph is \textit{spanning} (i.e. contains all the nodes of $G$), is a \textit{tree} (i.e. is connected and contains no cycles) and is \textit{minimal} in that the sum of the edge weights is minimal among all spanning trees. Every connected graph has a (essentially unique) MST.

In this work we consider sets of points in various spaces, with distance between points defined to be:
\begin{itemize}
	\item In two and three dimensions: Euclidean distance;
	\item On the sphere (i.e. RA, Dec.): subtended angle;
	\item Using RA, Dec., redshift: convert redshift to comoving distance (using the fiducial cosmology), then use Euclidean distance.
\end{itemize}
Given a set of points $S$ we wish to investigate the MST of the complete graph on these points (i.e. there is an edge between every pair of points and all these edges are candidates for inclusion in the MST); we refer to this as the MST of $S$. See Figure \ref{fig:mst_example} for an example of such an MST. Now Kruskal's algorithm \citep{kruskal1956shortest} (described below) takes as input a connected graph (not necessarily complete) and discards certain edges so as to find its MST. In theory, we should input to this algorithm the complete graph on $S$. However this is inefficient as the complete graph contains many edges (e.g. between widely separated points) that are very unlikely to appear in the output MST; it is sufficient to input to Kruskal's algorithm a pruned graph that retains only shorter edges.

To this end, we use as input to Kruskal's algorithm the $k$ nearest neighbours graph ($k$NN), i.e. the graph in which each point has an edge to its $k$ nearest neighbours. Here $k$ is a free parameter (and should not be confused with the wavenumber used in harmonic analysis). We calculate this graph using the \texttt{kneighbours\_graph} function from \texttt{scikit-learn}.\footnote{\href{http://www.scikit-learn.org}{http://www.scikit-learn.org}} Note that if $k$ is too small then the $k$NN graph need not be connected (it might consist of several isolated islands); in most cases considered, $k >10$ ensures that $k$NN will be connected (but when applying scale cuts (see Section \ref{scalecut}) a larger $k$ is needed).

We then apply the \texttt{scipy} \texttt{minimum\_spanning\_tree}\footnote{\href{https://scipy.org/}{https://scipy.org/}} function, which implements Kruskal's algorithm. This algorithm removes all the edges from the graph, sorts these removed edges by length (shortest to longest), and then sequentially re-inserts them, omitting an edge if its inclusion would create a cycle. This continues until all points are connected into a single tree. The Kruskal algorithm can be shown to scale as $\mathcal{O}(N_{\rm E}\log N_{\rm V})$ \citep[see][section on Kruskal's algorithm]{introtoalgorithms2009} where $N_{E}$ is the number of edges in the supplied spanning graph and $N_{\rm V}$ is the number of nodes. At most $N_{\rm E} \simeq N_{\rm V}^{2}$ but this can be greatly reduced by using the $k$NN graph, which changes the scaling from $\mathcal{O}(n^{2}\log n)$, where $n$ is the number of nodes, to $\mathcal{O}(kn\log n)$. Since usually $k\ll n$ this greatly reduces computation time.

We tested the sensitivity to the choice of $k$ by using a graph with $256^{3}$ points (HZ = High $\sigma_{8}$ and zero $\sum m_{\nu}$ simulations at $z=0$ explained later in Section \ref{section_nbody}). We compared the total length of the MST when $k=50$ (a proxy for $k=\infty$) and found a fractional difference of $\sim 2\times 10^{-6}$ for $k=20$, $\sim 2\times 10^{-7}$ for $k=30$, and $\sim 3\times 10^{-8}$ for $k=40$. It appears that $k=20$ gives a good balance between computation time and an accurate estimation of the MST, so we use this value except where stated otherwise.

\subsection{Statistics from the minimum spanning tree}

Any given MST is a complex structure with many interesting features. In this study, we are not interested in these individual features but rather the overall properties and their relation to cosmological parameters. Taking inspiration from \citet{Rainbolt2016} and \citet{Krzewina1996} we measure the probability distribution (i.e. histograms) of the following:
\begin{itemize}
	\item Degree ($d$): the number of edges attached to each node.
	\item Edge lengths ($l$): the length of edges.
	\item From branches, which are chains of edges connected with intermediary nodes of $d = 2$, we measure:
	\begin{itemize}
		\item Branch lengths ($b$): the sum of edges that make up the branch.
		\item Branch shape ($s$): the straight line distance between the branch ends divided by the branch length.
	\end{itemize}
\end{itemize}

These statistics are displayed in Figure \ref{fig:mst_example}. Of course one could consider other statistics to extract from the MST \citep[see][]{GAMA2014} but we choose to explore these as they have been shown to successfully aid in the classification of particle physics interactions \citep[see][]{Rainbolt2016}. The MST will have a total of $n-1$ edges \citep{kruskal1956shortest}, where $n$ is the number of nodes. Since each edge has a node on either end, each edge contributes twice to the total degree of the MST. Hence the expectation value for $d$ will be:
\begin{equation}
\label{eq:degree_mean}
\langle d \rangle = \frac{2 (n-1)}{n} \simeq 2.
\end{equation}
By definition the branch shapes satisfies $0 \leq s\leq 1$. Often $s$ is near $1$, so to facilitate visual comparison we frequently plot $\sqrt{1-s}$ instead of $s$. Straighter branches correspond to $\sqrt{1-s}$ closer to zero.

Additionally it is useful in certain circumstances, particularly when comparing MSTs that contain different number of nodes, to look at the dimensionless parameters of:
\begin{itemize}
	\item $\ln \left(\bar{l}\right)$, where $\bar{l}=l/\langle l\rangle$ and $\langle l\rangle$ is the average edge length.
	\item $\ln \left(\bar{b}\right)$, where $\bar{b}=b/\langle b\rangle$ and $\langle b\rangle$ is the average branch length.
\end{itemize}
Comparing the distribution of these dimensionless parameters is only appropriate if the distribution of points is scale-independent. In cosmology this is not necessarily the case for higher order statistics, so these should be used sparingly.

\begin{table*}
	\centering
	\caption{A summary of the simulation suites used in this study. For each simulation suite we list its name, the method used to produce it, the point distribution used and the use to which it is put.}
	\begin{tabularx}{\linewidth}{llll}
		\hline\hline
		\TBstrut Name & Method & Points & Usage\\
		\hline
		\Tstrut Illustris & Hydrodynamic  & Subhaloes & Testing the sensitivity of the MST to higher order statistics\\
		& & & (i.e. beyond two-point)\\
		MICE & $N$-body & Galaxies & Exploring the sensitivity to RSDs\\
		$\nu N$-body & $N$-body & Dark matter particles & Using an unbiased tracer we look to find what the MST\\
		& & and haloes & is actually measuring\\
		\Bstrut PICOLA & COLA & Haloes & Comparing sensitivity of the MST to traditional methods\\
		\hline\hline
	\end{tabularx}
	\label{simulation_summary}
\end{table*}

\subsubsection{Computational issues for finding branches}

Once the MST is constructed, we know the edge lengths ($l$) and the indices of the nodes at either end of the edges. These can be trivially used to find the degree ($d$) of each node and edge end. To find branches, we search for edges joining a $d=2$ node to a $d\neq2$ node (i.e. `branch ends') and edges joining two $d=2$ nodes (such edges, which form the middle parts of branches, are referred to as `branch mids'). To find the branches we begin with a branch end, search for a branch mid that is connected to it, and continue to grow the branch until no more branch mids can be added. At this point we then search for the branch end that finishes it. This is a computationally expensive procedure but can be trivially made faster by dividing the entire tree into sections and running the algorithm on the sections independently. Branches straddling the boundaries will be left incomplete, but can be completed by matching any remaining incomplete branches.

{\sc MiSTree} \citep{mistree}, the {\sc Python} package to construct the MST and derive its statistics, is made publicly available.\footnote{\href{https://github.com/knaidoo29/mistree}{https://github.com/knaidoo29/mistree}}

\subsection{Error estimation}
\label{error}

Uncertainties for the MST statistics are generated in two ways. 
\begin{itemize}
	\item In the cases where many realizations of a data set can be generated easily we will estimate the mean and standard deviation from an ensemble of realizations.
	\item If only a single realization is available we will use jackknife errors. Here, we divide up our data set into $n$ regions and run the analysis $n$ times, each time removing a single different region from the analysis yielding an output $\theta_{\rm i}$. The errors, $\Delta \theta_{\rm jack}$, are estimated using
	\begin{equation}
	\Delta\theta_{\rm jack} = \left[\frac{n-1}{n}\sum^{n}_{i=1} \left(\theta_{\rm i}-\bar{\theta}\right)^{2}\right]^{1/2},
	\end{equation}
	where $\bar{\theta}$ is the average of $\theta_{\rm i}$. 
\end{itemize}

\subsection{Simulation summary}
\label{sim_explanation}

We use several simulations suites; these are summarised in Table \ref{simulation_summary}. We discuss these simulations in greater detail in the relevant sections of the paper where they are used.

%% file: sensitivity.tex
\begin{figure*}
	\includegraphics[width=\textwidth]{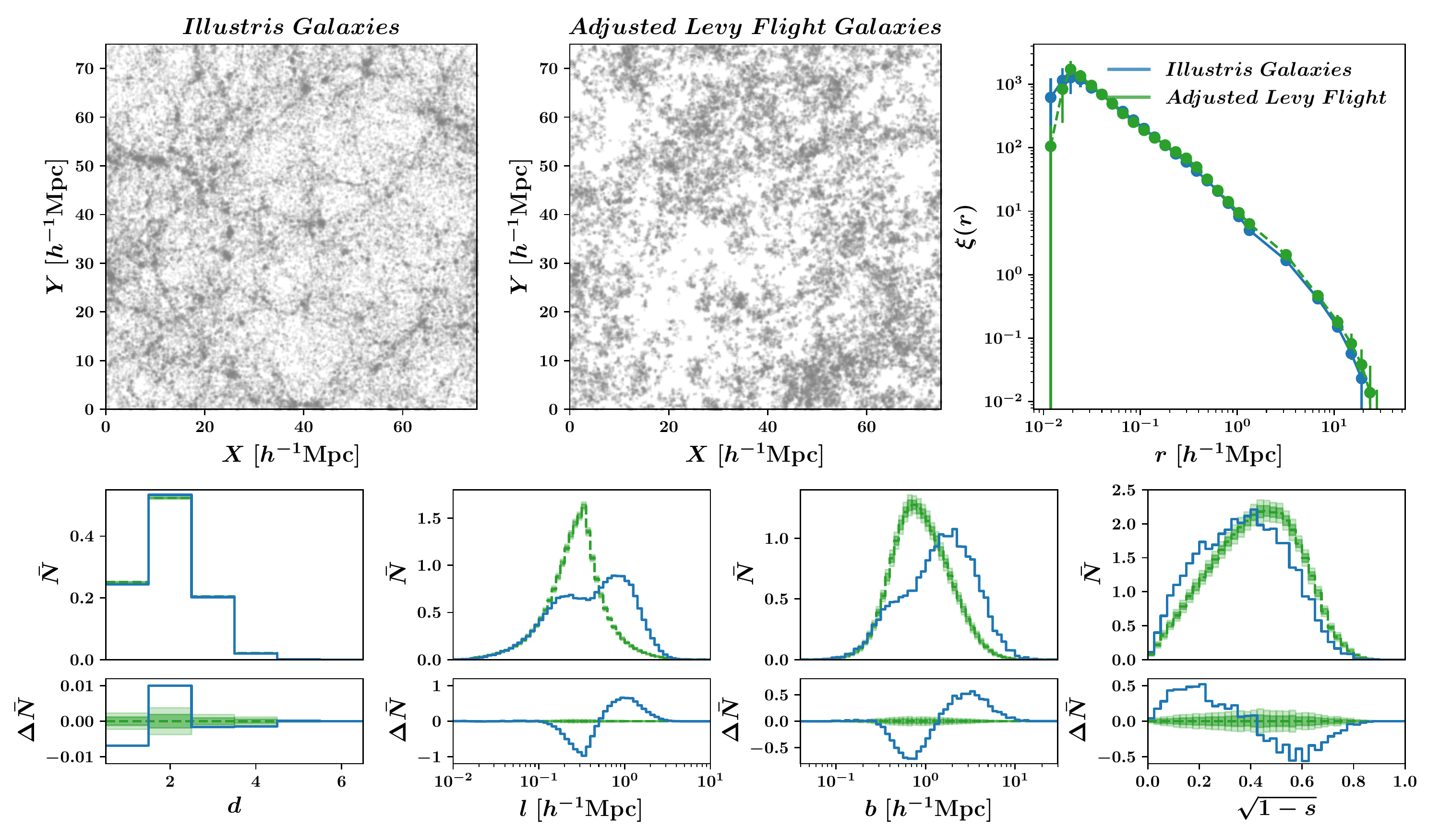}
	\caption{Top panels: the left shows the Illustris galaxy sample and the middle panel shows one realization of the ALF. Visually these two simulations are different in their distribution of galaxies. However they have virtually identical 2PCF by construction (right-hand panel). Illustris measurements are shown in blue and the mean for 100 realizations of the ALF is shown by the green dashed line and green envelopes show the 1$\sigma$ (darker) and 2$\sigma$ regions. Bottom panels: the histogram distributions of the MST statistics (from left to right): degree ($d$), edge length ($l$), branch length ($b$), and branch shape ($s$; note we plot the $\sqrt{1-s}$ value instead because the distribution peaks towards 1 and it is easier to see the difference in this projection). The difference between the PDF is displayed in the bottom subplots where zero on the $y$-axis corresponds to the mean counts for the ALF PDF. The measurements from the MST are significantly different for each of these simulations. In particular the distributions of edge lengths and branches show some bimodality for the Illustris sample which is not present in the ALF. This demonstrates the sensitivity of the MST to patterns in the cosmic web as the bimodal distribution appears to be driven by void and cluster environments (explored in Section \ref{mst_vs_density}).}
	\label{fig:levy_vs_illustris}
\end{figure*}

\section{Sensitivity of MST to Cosmic Web patterns}
\label{higherorder}

\subsection{Heuristic argument}
\label{heuristic}

There are compelling reasons to believe the MST should be sensitive to cosmic web patterns. Consider how the Kruskal algorithm constructs the MST (see Section \ref{method}). An edge is added only if this does not create a cycle; this means that the very construction of the MST requires an awareness of neighbouring edges or more generally the environment each edge inhabits. More generally this means the inclusion of a single edge is not defined solely by the 2PCF but by its local environment. Therefore, we should expect the MST to contain more information than is present in the 2PCF.

\subsection{Illustris vs. adjusted L\'{e}vy flight}

Testing whether the MST is sensitive to higher order statistics is rather challenging since at present there are no analytical descriptions of the MST statistics.

To go around this theoretical limitation we instead carry out an analysis similar to that of \citet{Hong2016}, comparing the Illustris\footnote{\href{http://www.illustris-project.org}{http://www.illustris-project.org}} \citep{Nelson2015,Vogelsberger2014} simulations (see Section \ref{illustris}) to an adjusted L\'{e}vy flight (ALF) simulation that is tuned to have almost identical 2PCF but different higher order information.

L\'{e}vy flights \citep{Mandelbrot1982} are random walk simulations where the step size (the distance between one point and the next) is given by a fat-tailed power-law probability distribution function (PDF). This ensures that its 2PCF will follow a power law \citep[see][]{Mandelbrot1982} similar to that found for galaxies. However, although a standard L\'{e}vy flight scheme may be able to replicate the 2PCF at large scales, at small scales, the 2PCF eventually plateaus \citep[see][]{Hong2016}. Since the MST is sensitive to small scales, it is important that the L\'{e}vy flight simulation match that of the Illustris sample at small scales. We are able to match the 2PCF of the Illustris sample at all scales using an adjusted L\'{e}vy flight (ALF) simulation as explained below.

\subsubsection{Illustris galaxy sample}
\label{illustris}
We use the subhalo catalogue of the Illustris-1 snap 100 sample and follow \citet{Hong2016} to include only subhaloes which are large and dark-matter-dominated:
\begin{equation}
\begin{array}{l}
M_{\rm *} \geq 10^{8} \si{\solarmass},\\
M_{\rm *} < 0.63M_{\rm DM},
\end{array}
\end{equation}
where $M_{\rm *}$ and $M_{\rm DM}$ are the stellar and dark matter mass of the subhaloes respectively. We will refer to this as the Illustris galaxy sample.

\subsubsection{Adjusted L\'{e}vy flight}

We generate an ALF simulation with the same number of `galaxies' as our Illustris sample and (almost) the same 2PCF. For comparison with Illustris we enforce periodic boundary conditions. The standard L\'{e}vy flight has step sizes $t$ with cumulative distribution function (CDF),
\begin{equation}
{\rm CDF}(t) = \left\{ \begin{array}{lcl}
0  & \mbox{for} & t < t_{\rm 0},\\
1 - \left(\frac{t}{t_{\rm 0}}\right)^{-\alpha} & \mbox{for} & t\geq t_{\rm 0},
\end{array} \right.
\end{equation}
where $t_{\rm 0}$ and $\alpha$ are free parameters. This yields a simulation with a power-law 2PCF of the form $C(t_{\rm 0},\alpha) t^{3-\alpha}$ at scales larger than $t_{\rm 0}$ (where $C(t_{\rm 0}, \alpha)$ is a constant determined by the free parameters), below this scale the 2PCF plateaus \citep[see][]{Hong2016}. To have control of the 2PCF below scales of $t_{\rm 0}$ we introduce an ALF model with the following CDF:
\begin{equation}
{\rm CDF}(t) = \left\{ \begin{array}{lcl}
0  & \mbox{for} & t < t_{\rm s},\\
\beta\left(\frac{t-t_{\rm s}}{t_{\rm 0}-t_{\rm s}}\right)^{\gamma}& \mbox{for} & t_{\rm s} \leq t < t_{\rm 0},\\
(1-\beta)\left[1 - \left(\frac{t}{t_{\rm 0}}\right)^{-\alpha}\right]+\beta & \mbox{for} & t\geq t_{\rm 0}.
\end{array} \right.
\end{equation}
This introduces three new parameters: $\beta$, $\gamma$, and $t_{\rm s}$. Rather than having a step size probability distribution function (PDF) that jumps from zero to a maximum at $t_{\rm 0}$, the ALF is constructed to have a slow rise to the maximum at $t_{\rm 0}$. The second piece of the CDF describes a transfer function that operates between $t_{\rm s}$ and $t_{\rm 0}$ (where by definition $t_{\rm s} < t_{\rm 0}$). Here $\gamma$ allows us to control the gradient of this rise and $\beta$ allows us to define the fraction of step sizes below $t_{\rm 0}$.

\begin{figure*}
	\includegraphics[width=\textwidth]{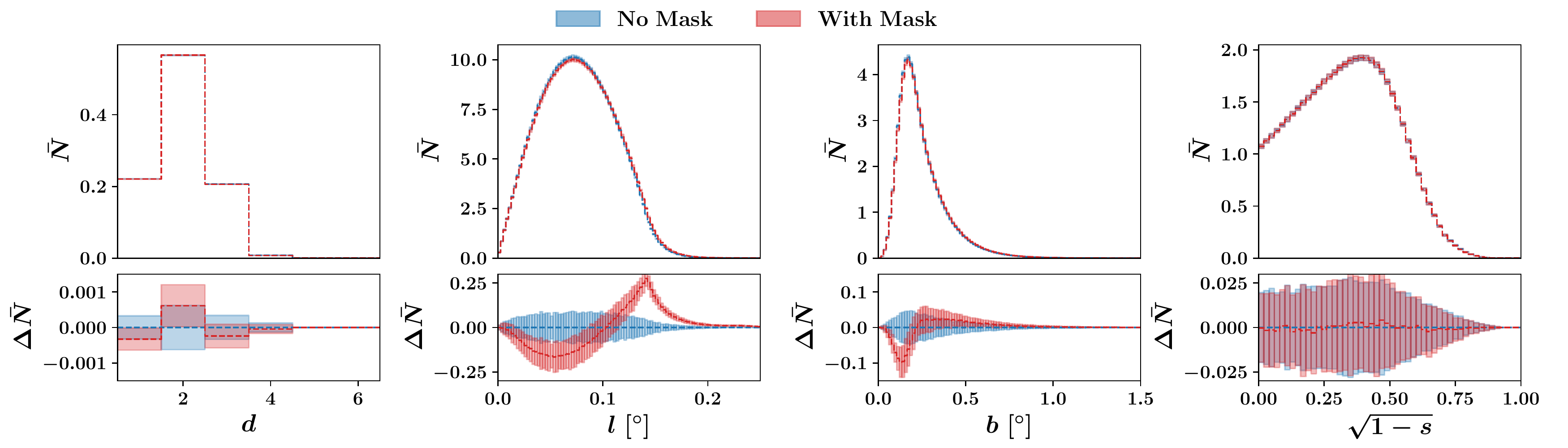}
	\caption{The MST statistics, calculated tomographically on random points placed in the BOSS CMASS North footprint (placed with the same density as the BOSS CMASS galaxies), with (red) and without (blue) using the CMASS mask. We see a significant shift towards longer edges in the MST performed with the mask, with a similar effect seen in the distribution of branch lengths. For the degree and branch shape the masking has no statistically significant effect.}
	\label{fig:realistic_mask}
\end{figure*}

\subsubsection{Comparison}

The Illustris sample contains \num{63453} galaxies. We create a sample of the same size using an ALF model with parameters $\alpha = 1.5$, $t_{0}=0.325$, $t_{\rm s}=0.015$, $\beta=0.45$, and $\gamma=1.3$ (where length-scales $t_{\rm 0}$ and $t_{\rm s}$ are given in $h^{-1}{\rm Mpc}$). The two samples have approximately equal 2PCFs down to scales of $0.01\ h^{-1}{\rm Mpc}$ by construction. The 2PCF was calculated on a single realization of the ALF model with varying $\beta$, $\gamma$, $t_{\rm s}$ and $t_{\rm 0}$ ($\alpha = 1.5$ was kept constant, see \citealt{Hong2016}). We then chose the parameters that produced the closest match, i.e. by minimizing the sum of difference between the 2PCF in log space. The Illustris and ALF sample show widely different MST statistics (see Figure \ref{fig:levy_vs_illustris}), thereby demonstrating the sensitivity of the MST to higher order statistics. The bimodal distribution of edge and branch lengths shown in Figure \ref{fig:levy_vs_illustris} occurs in over- and underdensities (explored in more detail in Section \ref{section_nbody}). Note also that we see differences in the shape of branches and the distribution of degrees to a statistically significant level, although these differences are not as striking as the difference in edge and branch length distributions.

%% file: systematics.tex
\section{Boundary effects and Redshift Space Distortions}
\label{systematics}

We study possible sources of systematic errors that could affect the MST. In particular we would like to establish to what extent simulations need to replicate survey properties.

\subsection{Boundary effects}
\label{Boundary}

Galaxy surveys often contain complex survey footprints with regions masked due to stars and varying completeness and it is important to understand how such footprints will affect the MST. Imposing a mask on the data set results in two effects:
\begin{enumerate}
	\item Additional edges are included to join nodes near the boundaries. These would have otherwise been joined by nodes outside the boundary in a larger MST.
	\item New edges are located near the centre whose purpose appears to be to unify the structure as a single spanning tree. In a larger spanning tree, these separated regions would be connected through routes that extend beyond the boundary.
\end{enumerate}
The net result of these effects is to create a slight bias towards longer edges and slightly longer branches. Interestingly, all edges in the larger MST (within the boundary) are present in the smaller MST. This property always holds, as can easily be proven using the `cycle property' of the MST \citep[see][]{katriel2003practical}.

We investigate the effects of a realistic mask by using the BOSS CMASS MD-Patchy mocks North mask \citep{bosscmass2016}, which includes masking for bright stars, bad fields, centrepost and collision priority.\footnote{See \href{http://www.sdss3.org/dr9/algorithms/boss_tiling.php\#veto_masks}{http://www.sdss3.org/dr9/algorithms/boss_tiling.php\#veto_masks}} In Figure \ref{fig:realistic_mask} we demonstrate the effects of this mask on random points placed within the CMASS footprint (with the same density as the CMASS galaxies) with and without a mask. The MST is then calculated on 1000 realizations tomographically (i.e. on the sphere). The degree and branch shape show little change but the distribution of edge lengths show a significant tendency towards longer edges when a mask is used. This is mirrored by a similar effect in the distribution of branch lengths. This is because the mask eliminates shorter paths, forcing the MST to include longer edges that would (without the mask) have been excluded. This demonstrates that realistic masks with holes do have an impact on the MST and must be included in any future analysis.

\subsection{Scale cuts}
\label{scalecut}

In cosmology there is often a need to apply scale cuts in real space. This can occur for a variety of reasons: theoretical uncertainty at small scales both from simulation and from analytic formulae and also practically from fibre collisions in spectroscopic surveys. For the 2PCF this is rather simple to mitigate; you simply restrict the domain of the 2PCF to exclude separations below the scale cut. With the MST this is more complicated. Unfortunately there does not appear to be a way to deal with this after the MST has been constructed; this is because the problematic smallest edges will by construction be incorporated in the graph. To ensure that problematic small scales are removed from the MST we alter the $k$NN graph that is the input to the Kruskal algorithm by removing edges whose length is below the desired scale cut.

\begin{figure*}
	\includegraphics[width=\textwidth]{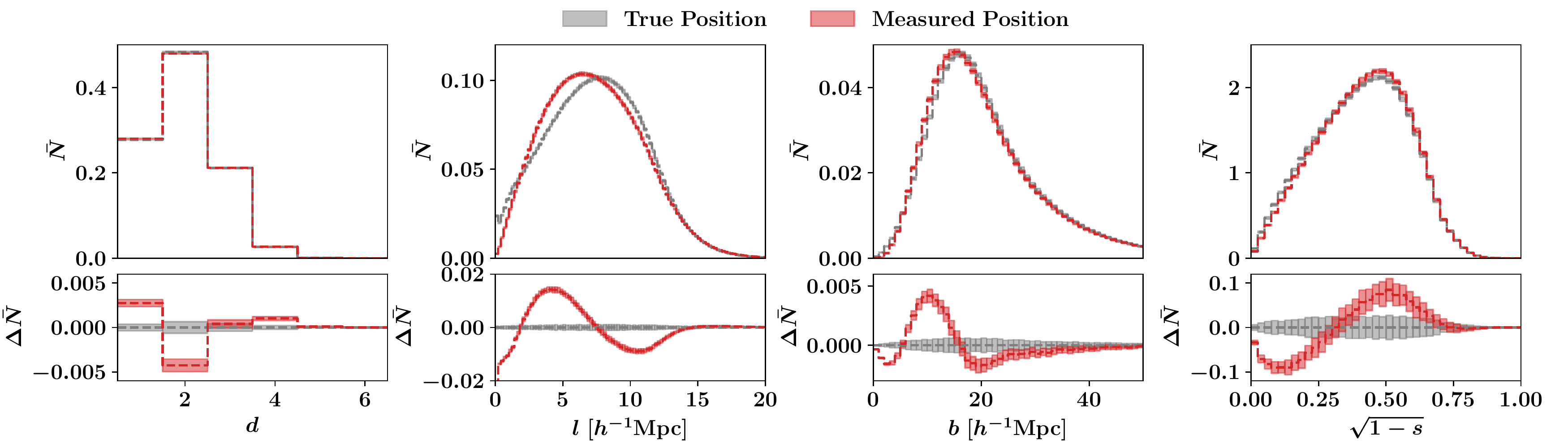}
	\caption{The effects of RSDs on the MST statistics. From left to right: the MST statistics degree ($d$), edge length ($l$), branch length ($b$), and branch shape ($s$). Bottom panels show the differences. Ten realizations of \num{500000} MICE galaxies were generated and the MST were constructed on their true positions (grey) and then the measured positions (red), i.e. the inferred positions based on their redshifts including RSD. The envelopes correspond to $1\sigma$ uncertainties. Significant differences between the MST statistics show that the MST is sensitive to the RSD effect.}
	\label{fig:rsd_effect_mst}
\end{figure*}

\subsection{Redshift space distortion on MICE galaxies}

RSDs \citep{Kaiser1987}, caused by the Kaiser and Fingers of God effects, will distort the measured redshift of galaxies and thus will impact the inferred comoving distance. Since this effect alters the 3D distribution of galaxies, it will inevitably affect the MST statistics.

We explore this effect by comparing the MST performed on a subset of the MICE galaxy catalogue \citep{MICE2015} in real and redshift space (i.e. with RSD). Here, we randomly draw \num{10} realizations of \num{500000} galaxies with real comoving distances between \num{1000} to \num{1500} $h^{-1}{\rm Mpc}$. We ensure that the density of galaxies is constant so that the number of galaxies $\propto D_{\rm c}^{3}$, where $D_{\rm c}$ is the radial comoving distance from the observer.

Figure \ref{fig:rsd_effect_mst} shows the MST statistics with and without the RSD effect.  We see significant results in all the MST statistics demonstrating the importance of including this effect in any future MST study.

%% file: nbody.tex
\section{What does the minimum spanning tree measure?}
\label{section_nbody}

This section considers the following questions: 
\begin{enumerate}
	\item What do the MST statistics look like on an unbiased tracer (i.e. $N$-body dark matter particles)?
	\item What does the MST statistics tell us about the underlining density distribution?
	\item What is the relation of MST statistics to 2PCF?
	\item What happens when we change simulation resolution?
	\item How do the MST statistics change when measured on haloes (i.e. a more galaxy-like tracer)?
\end{enumerate}

\begin{table*}
	\centering
	\caption{Simulation and cosmological parameters for the $N$-body simulations. \citet{Massara2015} uses different names, which we list here.}
	\begin{tabularx}{\linewidth}{llllccccc}
		\hline\hline
		\TBstrut Name & Reason for name & \citet{Massara2015} & $L_{\rm Box}$ ($h^{-1}{\rm Mpc}$) & $N_{\rm cdm}$ & $N_{\nu}$ & $\sum m_{\nu}$ (\si{\electronvolt}) & $\sigma_{8}$ & $10^{9}A_{\rm s}$\\
		\hline
		\Tstrut HZ & High $\sigma_{8}$, zero $\sum m_{\nu}$ & L0 & $1000$ & $256^{3}$ & $0$ & $0$ & $0.834$ & $2.13$\\
		LZ & Low $\sigma_{8}$, zero $\sum m_{\nu}$  & L0s8 & $1000$ & $256^{3}$ & $0$ & $0$ & $0.693$ & $1.473$\\
		LN & Low $\sigma_{8}$, non-zero $\sum m_{\nu}$  & L60 & $1000$ & $256^{3}$ & $256^{3}$ & $0.6$ & $0.693$ & $2.13$\\
		HZHR & High $\sigma_{8}$, zero $\sum m_{\nu}$, high resolution & H0 & $500$ & $512^{3}$ & $0$ & $0$ & $0.834$ & $2.13$\\
		\Bstrut LNHR & Low $\sigma_{8}$, zero $\sum m_{\nu}$, high resolution & H60 & $500$ & $512^{3}$ & $512^{3}$ & $0.6$ & $0.693$ & $1.473$\\
		\hline\hline
	\end{tabularx}
	\label{tableparam}
\end{table*}

\subsection{$\nu N$-body simulations}

Five $N$-body simulations \citep[see][]{Massara2015} were made by running the {\sc TreePM} code {\sc GADGET-III} \citep{gadget2}. The following cosmological parameters were common to all simulations: $\Omega_{\rm m}=0.3175$, $\Omega_{\rm b}=0.049$, $\Omega_{\Lambda}=0.6825$, $h=0.6711$, and $n_{\rm s}=0.9624$. See table \ref{tableparam} for a list of the simulations used and their respective cosmological parameters, particle numbers and box sizes. The cold dark matter energy density is set to $\Omega_{\rm c} = \Omega_{\rm m}-\Omega_{\rm b}-\Omega_{\nu}$ where $\Omega_{\nu}h^{2}\simeq\sum m_{\nu}/(94.1\rm\ \si{\electronvolt})$. Cold dark matter and neutrinos are both treated as collisionless particles. They differ in their masses and in their initial conditions, where the initial conditions for neutrinos receive an extra thermal velocity obtained by randomly sampling the neutrino Fermi--Dirac momentum distribution \citep{Viel2010}. These are evolved from an initial redshift of $z=100$. Table \ref{tableparam} summarised the simulations used.

\subsection{MST application to dark matter particles}

\begin{figure*}
	\includegraphics[width=\textwidth]{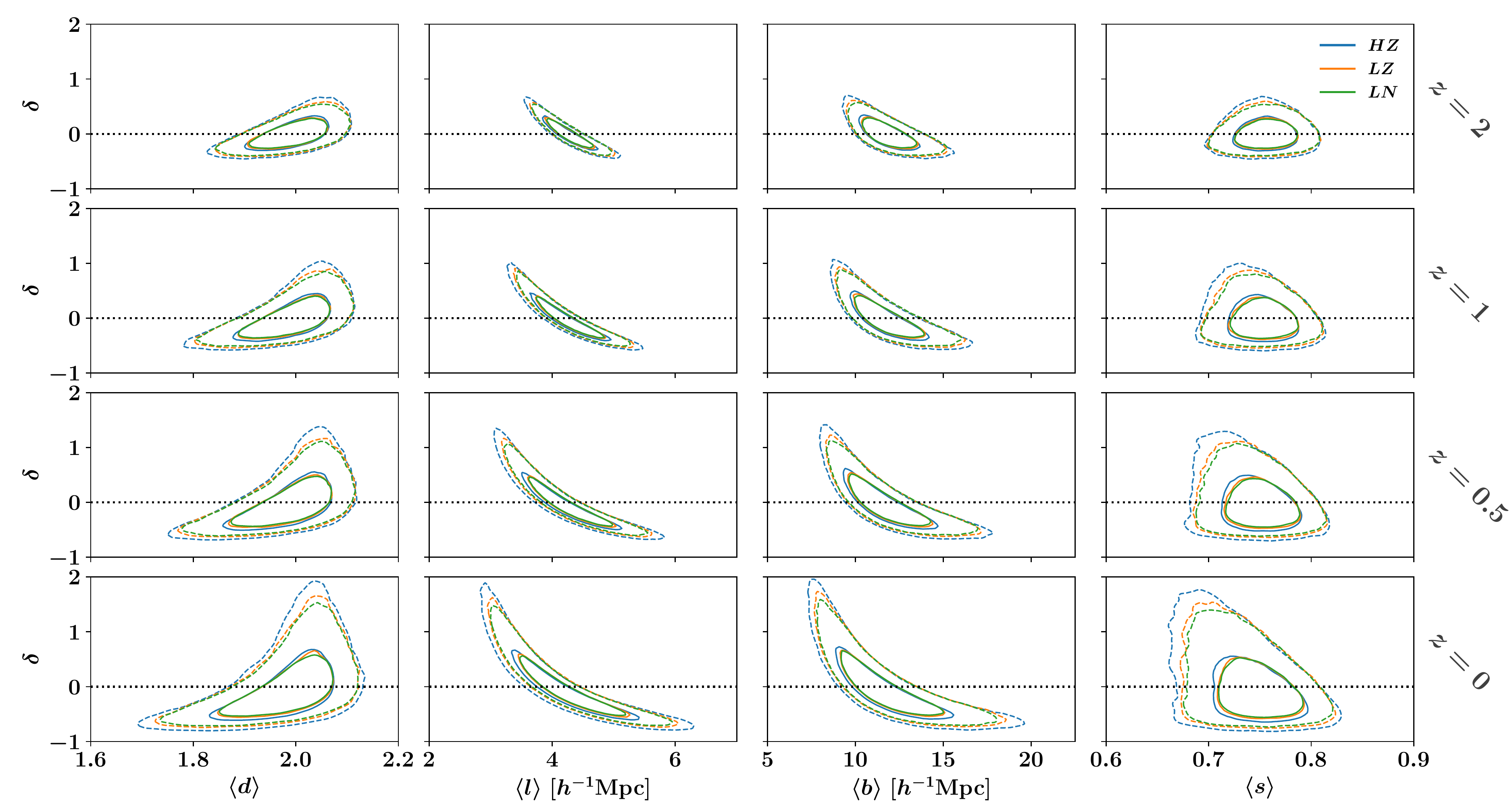}
	\caption{Contour plots of the average density contrast ($\delta$) is plotted against the MST statistics [from left to right: the average degree ($\langle d \rangle$), edge length ($\langle l\rangle$), branch length ($\langle b\rangle$) and branch shape ($\langle s \rangle$)] in $25\ h^{-1}{\rm Mpc}$ cubes. The $1\sigma$ and $2\sigma$ contours are indicated by solid and dashed lines, respectively. The relation for HZ is in blue, LZ in orange and LN in green. See Section \ref{mst_vs_density} for a detailed explanation of the relation and their evolution.}
	\label{fig:nbody_mst_vs_density}
\end{figure*}

\begin{figure*}
	\includegraphics[width=\textwidth]{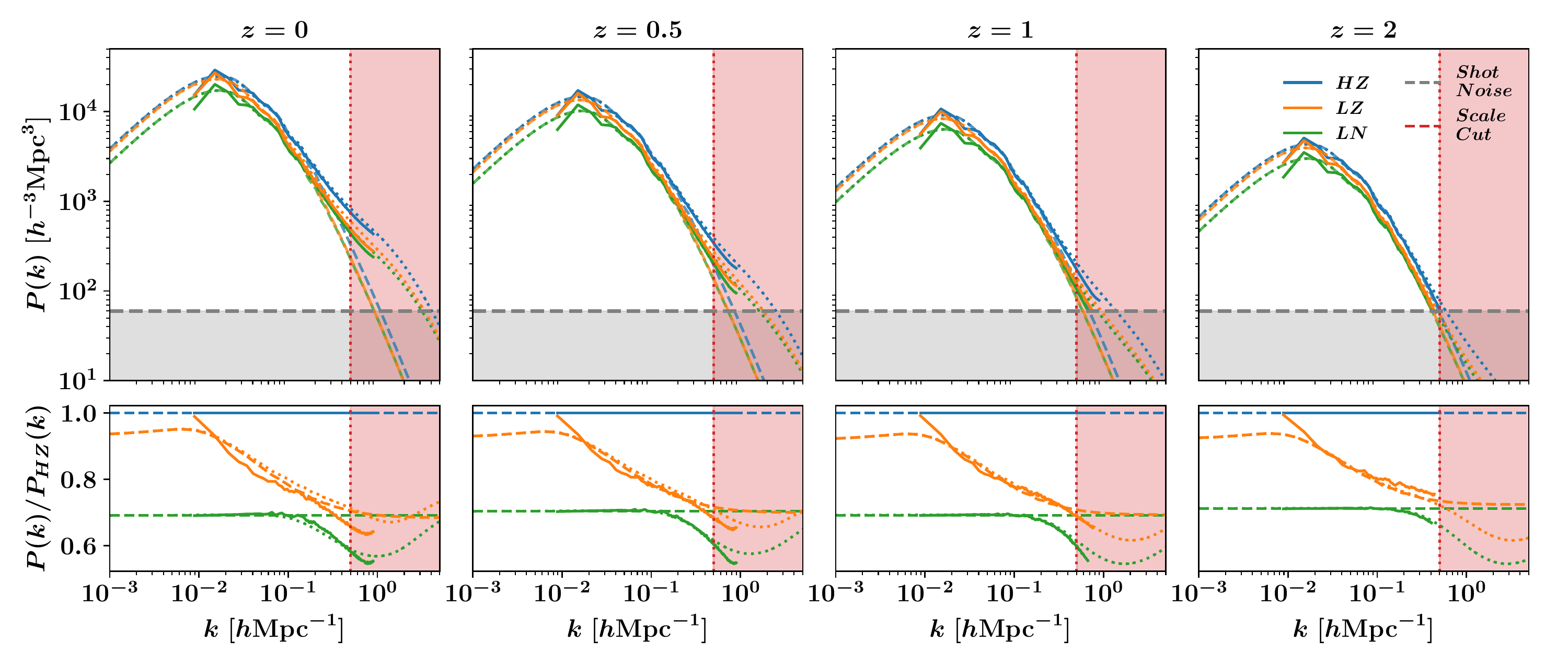}
	\caption{In the top panels the matter power spectra, $P(k)$, are plotted for redshift (from left to right) \num{2}, \num{1}, \num{0.5} and \num{0} for simulations HZ (blue), LZ (orange) and LN (green). In the bottom subplots we plot the ratio with respect to the HZ power spectra.  Solid lines correspond to the measured $P(k)$ from the respective simulations, while dashed and dotted lines correspond to the theoretical linear and  non-linear $P(k)$, respectively. The dashed grey lines shows the level at which the measured $P(k)$ will be affected by the shot noise of the simulation and the regions in red show the scales for which we apply a scale cut in the construction of the MST. Here, we see (that at all redshifts){\tiny } the power at all $k$ is highest for HZ, and then LZ and lastly LN. Note that LZ is close to HZ at high $k$ and close to LN at low $k$.}
	\label{fig:power_spec}
\end{figure*}

An MST was constructed on the dark matter particles from the HZ, LZ, and LN simulations (see table \ref{tableparam}), where errors were calculated using the jackknife method (Section \ref{error}). Figures \ref{fig:nbody_mst}, \ref{fig:nbody_mst_vs_density}, \ref{fig:power_spec}, and \ref{fig:haloes_mst} use the same colour scheme: HZ in blue, LZ in orange and LN in green. We boost the speed of the MST calculation by allowing this to be done in parallel, breaking the $N$-body snapshots into 64 cubes. We then implement the scale cut strategy discussed in Section \ref{scalecut} and partition the data set into four groups (to dilute the sample to look at larger sales) and apply a scale cut of $l_{\rm min}=2\ h^{-1}{\rm Mpc}$.

\subsubsection{Features in the minimum spanning tree statistics}
\label{dm_sims_mst}

\begin{figure*}
	\includegraphics[width=\textwidth]{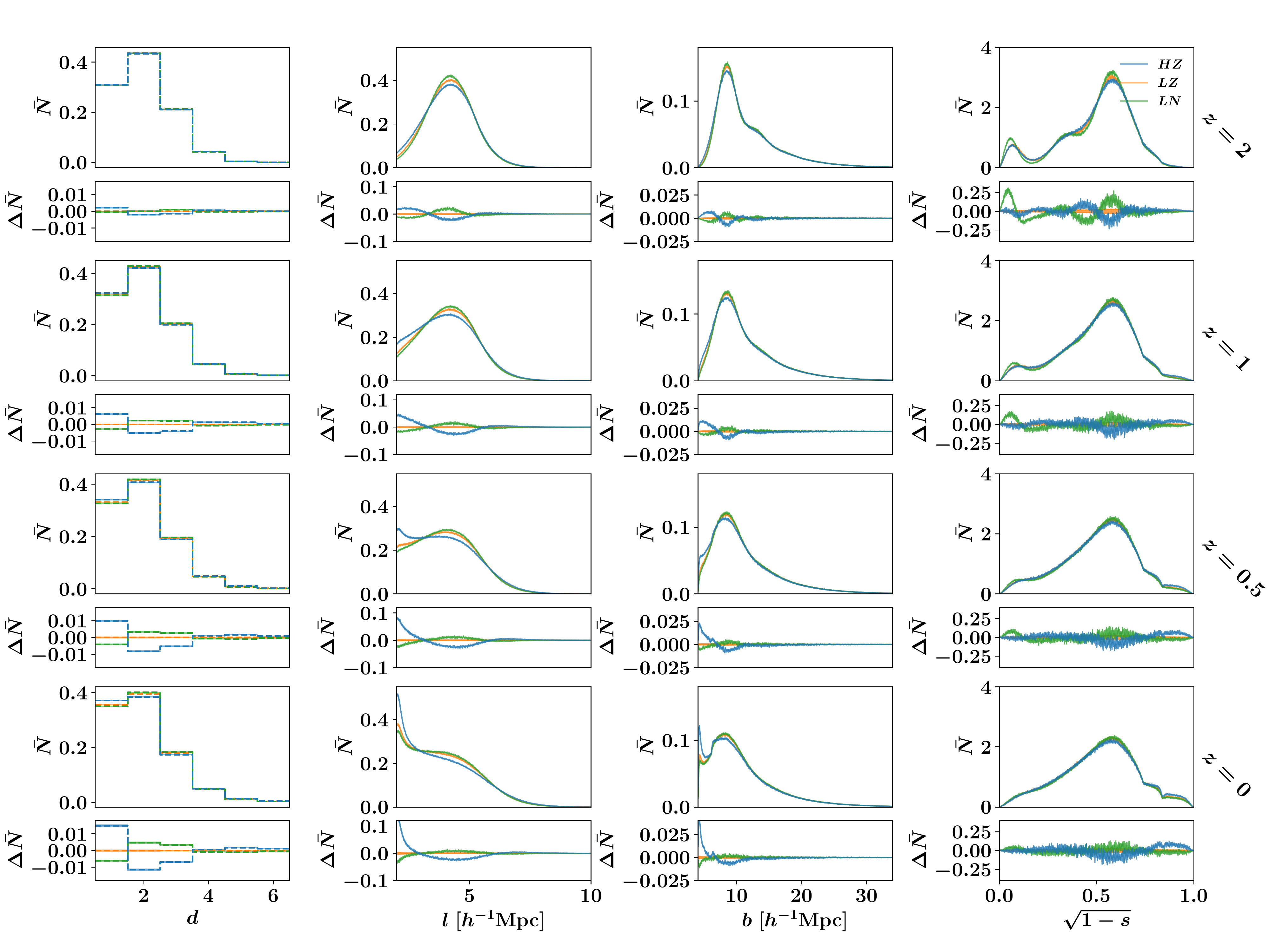}
	\caption{From left to right: the distribution of degree ($d$), edge length ($l$), branch length ($b$) and branch shape ($s$). These are obtained by dividing the full 1 $(h^{-1}{\rm Gpc})^{3}$ box into 250 $(h^{-1}{\rm Mpc})^{3}$ cubes for speed. These are then partitioned into four groups to minimise the effect of applying a scale cut of 2 $h^{-1}{\rm Mpc}$. From top to bottom: distributions are shown with respect to redshift \num{2}, \num{1}, \num{0.5} and \num{0}. These are further subdivided into a top subplot of the distributions and a bottom subplot of the differences. Simulations shown are HZ (blue), LZ (orange) and LN (green). See Section \ref{dm_sims_mst} for a detailed explanation of the distribution features, differences, and evolution.}
	\label{fig:nbody_mst}
\end{figure*}

In Figure \ref{fig:nbody_mst} we plot the MST statistics for these different simulations at redshifts $z=2$, $1$, $0.5$ and $0$. The plots display how the MST statistics evolve over cosmological time, as discussed below:
\begin{itemize}
	\item Degree: the distribution of degree remains relatively similar in all simulations and does not appear to evolve greatly over redshift, although differences between the simulations become more pronounced at lower redshifts.
	\item Edge length: overall we see that the distribution shows a high sensitivity to redshift, evolving from a single distribution into a bimodal one at smaller redshift.
	\begin{itemize}
		\item $l \geq 3\ h^{-1}{\rm Mpc}$: a broad peak is seen in the distribution at $l\simeq4\ h^{-1}{\rm Mpc}$. This feature dampens at lower redshift with the peak consistently highest for LN, followed by LZ and then HZ.
		\item $l < 3\ h^{-1}{\rm Mpc}$: a secondary peak emerges and dominates at lower redshift, which rises against the scale cut limit of $l_{\rm min}=2$.
		\item $l\sim3\ h^{-1}{\rm Mpc}$: between the two peak features is a region where seemingly all three distributions appear to converge and the orderings of the peaks above and below this point switch.
	\end{itemize}
	\item Branch length: the evolution appears virtually identical to the edge length distribution except at larger scales.
	\item Branch shape:
	\begin{itemize}
		\item A broad peak at $\sqrt{1-s} = 0.6$ which is present in all simulations. This peak is always highest for LN followed by LZ and HZ.
		\item A subpeak at $\sqrt{1-s}\sim 0.05$ which dampens at lower redshift. This suggests that some branches at low redshift are fairly straight. Since the simulation we use are fairly low in resolution we suspect that this feature is more an indication that the particles have not undergone much mixing and are still very close to their initial perturbed grid layout. This could be used as a diagnostic to test whether $N$-body simulations have moved from their perturbed gridded initial conditions.
		\item Lastly we see the emergence of two bumps between $\sqrt{1-s}\sim 0.7-1$ at low redshift. Comparison of the branch shape statistics with and without a scale cut show this is caused by the introduction of the scale cut, which forces some branches to be more curved. Branch shapes without a scale cut rarely see $\sqrt{1-s}>0.8$.
	\end{itemize}
\end{itemize}

\subsubsection{Exploring the minimum spanning tree relation to density}
\label{mst_vs_density}

To gain a greater physical intuition of what these statistics are telling us about cosmology, we subdivide the $1\ h^{-1}{\rm Gpc}$ cube into smaller $25\ h^{-1}{\rm Mpc}$ cubes. In these cubes we calculate the density contrast $\delta$,
\begin{equation}
\delta = \frac{N_{\rm DM}}{\langle N_{\rm DM}\rangle} - 1,
\end{equation}
where $N_{\rm DM}$ is the number of dark matter particles in a particular cube and and $\langle N_{\rm DM}\rangle$ is the average across all cubes. Figure \ref{fig:nbody_mst_vs_density} illustrates the relationship between the average degree ($\langle d\rangle$), edge length ($\langle l\rangle$), branch length ($\langle b\rangle$) and branch shape ($\langle s\rangle$) and the density contrast inside these cubes. 

\begin{itemize}
	\item $d$ vs $\delta$: we see that the mean of the degree, $d$, is relatively constant at $d\simeq2$ as a function of density. The variance shows a strong dependence on density,  with over densities having very low variance, i.e. predominantly $d=2$, and under densities showing a much larger variance and a slight tilt towards $d=1$. Of course we should expect high-density environment to form the main `backbone' of the MST, since these are the areas where the edges are shortest.
	\item $l$ and $b$ vs $\delta$: both the edge and branch length distribution show a very similar relation to density. Shorter edges and branches are mostly associated with overdensities and vice versa. Furthermore as the simulations evolve in redshift this relation becomes more pronounced. In both these statistics, we see that HZ appears consistently to have more overdense and underdense regions than the other two simulations. We also see that LN appears to have marginally but consistently higher overdense and underdense regions than LZ.
	\item $s$ vs $\delta$: the mean of the branch shape appears centred at 0.75 and shifts slightly to a mean of 0.7 for higher densities. Furthermore, as with the degree, the biggest relation to density is with the variance, which increases as the density lowers.
\end{itemize}
This analysis demonstrates a clear relation between MST statistics and environment (i.e. the local density).

\begin{figure*}
	\includegraphics[width=\textwidth]{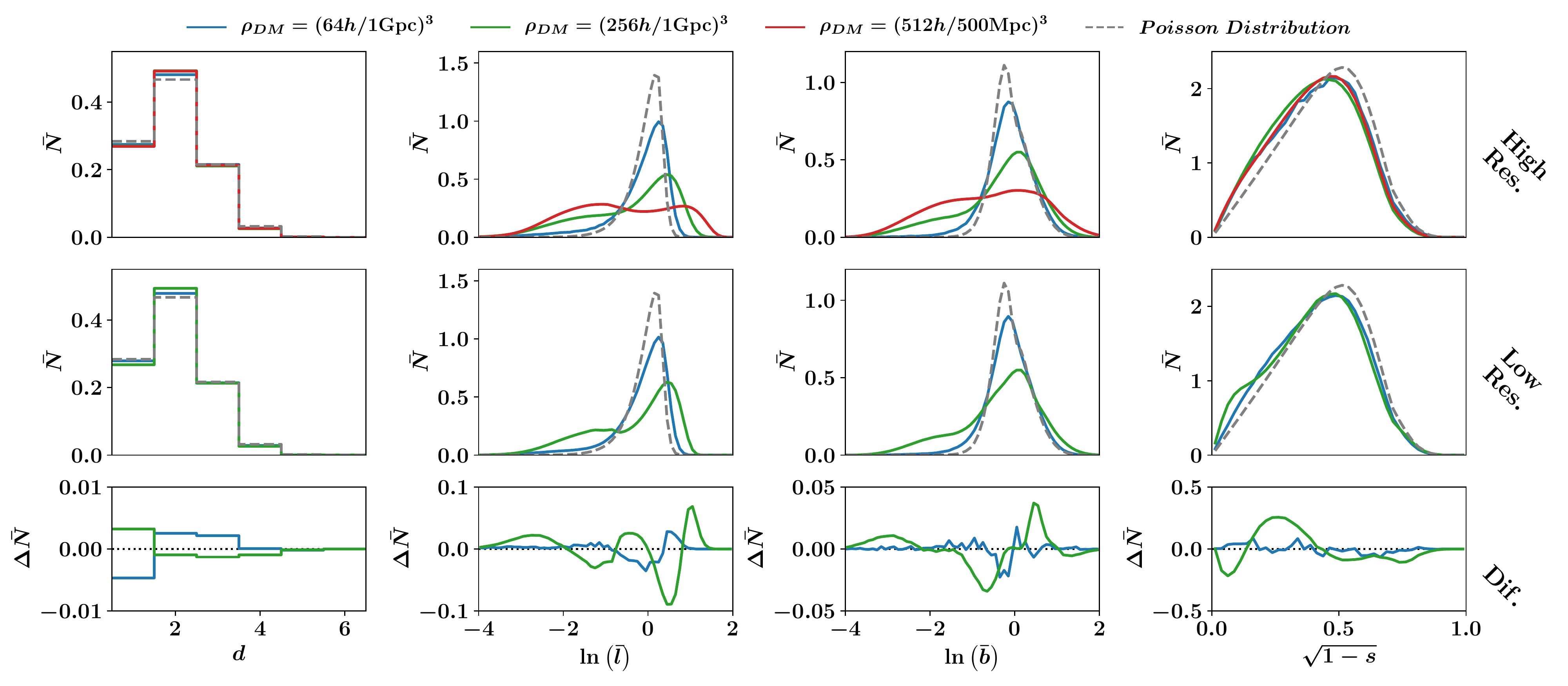}
	\caption{The distribution of degree ($d$), normalized edge and branch lengths ($\ln \left(\bar{l}\right)$ and $\ln \left(\bar{b}\right)$) and branch shape ($s$)  are displayed from left to right. Top panel: the distributions of the high-resolution versions of HZ (dashed line) and LN (dotted line) (i.e. HZHR and LNHR) simulation are shown in red and subsequently subsampled versions are shown in green and blue with dark matter particle densities ($\rho_{\rm DM}$) $256^{3}$, $128^{3}$ and $64^{3}$ per $(h^{-1}{\rm Gpc})^{3}$ respectively. Middle panel: the distribution for the HZ (dashed line) and LN (dotted line) simulation is shown. Bottom panel: the differences between the high resolution (top panel) and low-resolution (middle panel) simulations are shown. We additionally illustrate the distribution for random points (dashed grey).}
	\label{fig:resolution}
\end{figure*}

\begin{figure*}
	\includegraphics[width=\textwidth]{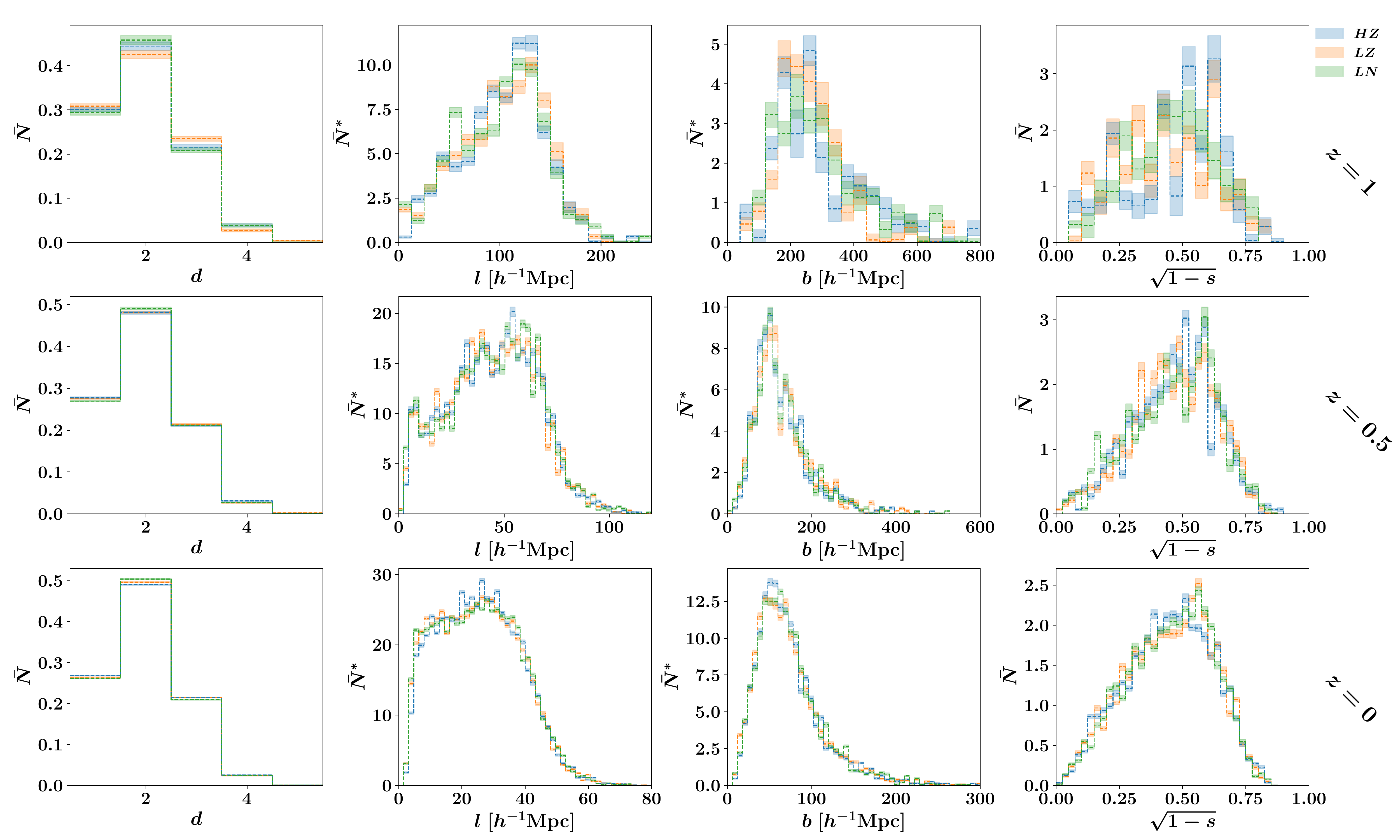}
	\caption{The MST constructed on halo catalogues derived from the HZ (blue), LZ (orange), and LN (green) $N$-body simulations. From left to right are the MST statistics: degree ($d$), edge length ($l$), branch length ($b$) and branch shape ($s$). They are plotted from top to bottom according to snapshots at redshift \num{1}, \num{0.5}, and \num{0}. Corresponding shaded areas show the jackknife uncertainties in the measurements. The distribution of the MST statistics are indistinguishable from each other at all redshifts, demonstrating that we should expect to see similar lines of degeneracy as power spectrum. Note $\bar{N}^{*}=10^{3}\bar{N}$.}
	\label{fig:haloes_mst}
\end{figure*}

\subsubsection{Relation to the matter power spectrum}

In Figure \ref{fig:power_spec}, we calculate the matter power spectra, $P(k)$, measured from these simulations. The dependence on redshift can be characterized by a simple shift in amplitude. We see that (at all $k$) HZ has more power, followed by LZ and then LN. At small $k$, LZ converges to HZ while at large $k$, LZ converges to LN. Notice, that the strength in $P(k)$ at large $k$ is matched by a tendency for shorter edges in the MST, demonstrating the MST expected dependence on clustering.

\subsubsection{Simulation resolution}
\label{resolution}

The MST of $N$-body simulations will be affected by the resolutions used. To measure the sensitivity of the MST statistics to the simulation resolution we calculate the MST on higher resolution versions of HZ and LN called HZHR and LNHR (see Table \ref{tableparam} for details of simulation properties). The resulting distributions of the MST statistics are shown in Figure \ref{fig:resolution}. For comparison we additionally subsample these two simulation boxes by randomly selecting particles in the simulation with equal number of particles. In the more sparsely sampled version of HZHR and LNHR the more resolved extreme high- and low-density environments are still imprinted. This can be seen by the fact that in the bottom panels of Figure \ref{fig:resolution} there appears to be more features at high and low values of $l$. This illustrates the importance of high-resolution simulations on the MST profiles inferred. We could also use highresolution simulations to calibrate the scale cut for low-resolution simulations by allowing the scale cut to vary until the MST statistics reach agreement between the high- and low-resolution simulations. We additionally measure the MST on a completely random set of points (shown in grey) illustrating how the more sparsely subsampled data set appears to be asymptotically approaching these profiles.

\subsection{MST application to haloes}

\begin{table}
	\centering
	\caption{The number of haloes found in each simulation (HZ, LN and LZ) for each redshift ($z$) snapshot. The number of haloes at $z=2$ was far too little for a meaningful MST and presumably would be uninformative.}
	\begin{tabularx}{0.55\columnwidth}{llll}
		\hline\hline
		\TBstrut Redshift & HZ & LN & LZ\\
		\hline
		\Tstrut0 & 17911 & 11168 & 9892\\
		0.5 & 6717 & 3017 & 2392\\
		1 & 1585 & 458 & 262\\
		\Bstrut2 & 16 & 2 & 1\\
		\hline\hline
	\end{tabularx}
	\label{table}
\end{table}

Halo catalogues were derived from the HZ, LZ, and LN simulation snapshots. We study these to get a sense of what the MST statistics will look like when performed on a biased tracer, such as galaxies. We dropped the $z=2$ snapshots as they contained too few haloes to be meaningful. Unlike the $N$-body simulation, we do not apply a scale cut since the density of haloes is quite low and the fraction of edges below $l_{\rm min}=2\ h^{-1}{\rm Mpc}$ is very low. The MST statistics derived from the haloes is shown in Figure \ref{fig:haloes_mst}. The number of haloes varies both across simulations and across redshift snapshots (see Table \ref{table}) - this is different from dark matter particles whose number count is constant across redshift and simulations.

To mitigate this issue, for each redshift we match the number of haloes to the lowest number found in the simulations (thus always matching the number of haloes found in the LZ simulations). For those with more haloes, we simply select the most massive haloes. In Figure \ref{fig:haloes_mst}, we find no real noticeable difference in the statistics suggesting the degeneracies of the MST may be similar to that found for $P(k)$.

%% file: comparison.tex
\section{Comparing the sensitivity to cosmology of power spectrum, bispectrum, and the minimum spanning tree}
\label{comparison}

In this section we compare the sensitivities to cosmological parameters of power spectrum $P(k)$, bispectrum $B(k_{1}, k_{2}, k_{3})$ and MST, measured on the same halo catalogues, to establish whether the MST can improve parameter constraints. Specifically, we compare the constraints on $A_{\rm s}$, $\Omega_{\rm m}$ and $\sum m_{\nu}$ for 10 sets of mock simulations. To obtain reliable posterior distributions for the three methods and their joint constraints, we would normally run an Markov Chain Monte Carlo (MCMC) using an analytic expression for the data vector. However, the MST statistics cannot be obtained analytically and hence have to be obtained from simulations. $P(k)$, $B(k_{1}, k_{2}, k_{3})$, and MST are therefore estimated from a grid of simulations in parameter space. To limit the noise in the estimates of the theory we take the mean of five simulations rather than just one at each point in parameter space. Additionally, since our simulation grid is rather sparse we use Gaussian process (GP) regression to interpolate the data vector. Finally we use a corrected likelihood function \citep[see][]{SellentinHeavens2016,Jeffrey2018} which accounts for the use of an estimated covariance matrix.

\subsection{COLA simulation suites}

A suite of COLA \citep{COLA} simulations were constructed using the {\sc MG-PICOLA} software \citep[][an extension to {\sc L-PICOLA} by \citealt{PICOLA}]{MG_PICOLA} which, among other things, can model the effects of massive neutrinos \citep{MG_PICOLA_Neutrinos}. This allowed us to generate $N$-body-like simulations relatively cheaply (in terms of computation time), albeit by sacrificing accuracy at small scales. All simulations are run in boxes of lengths 250 $h^{-1}{\rm Mpc}$, with $256^3$ dark matter particles and a discrete Fourier transform (DFT) density grid of $(3\times256)^3$. The latter is set to satisfy a requirement to produce accurate haloes from COLA simulations \citep{ice_cola}. The dependence on $A_{\rm s}$, $\Omega_{\rm m}$, and $\sum m_{\nu}$ are explored, while $h=0.6711$, $\Omega_{\rm b}=0.049$, and $n_{\rm s}=0.9624$ are constant in all simulations. Haloes and particles are outputted at redshift $z=0.5$, using 20 steps from an initial redshift $z=10$. Further details on the simulation suites are summarised in Table \ref{table:cola_simulation_suites}.

\begin{table*}
	\centering
	\begin{tabularx}{\linewidth}{lcccll}
		\hline\hline
		\TBstrut Name & $10^{9}A_{\rm s}$ & $\Omega_{\rm m}$ & $\sum m_{\nu}$ $[\si{\electronvolt}]$ & Realisations & Notes\\
		\hline
		\Tstrut	Grid & [1, 3.5] & [0.2, 0.5] & [0, 0.6] & 5 & Simulations carried out at $216$ points defined across a $6\times6\times6$ grid \\
		& & & & & in parameter space. \\
		Fiducial & 2 & 0.3 & 0 & 500 & Used to calculate covariance matrices.\\
		\Bstrut Mock & 2.13 & 0.3175 & 0.06 & 10 & Treated as real data.\\
		\hline\hline
	\end{tabularx}
	\caption{Properties of the simulations suites are shown above; including the reference names, cosmological parameters, realisations and information on their eventual uses.}
	\label{table:cola_simulation_suites}
\end{table*}

\begin{figure}
	\includegraphics[width=\columnwidth]{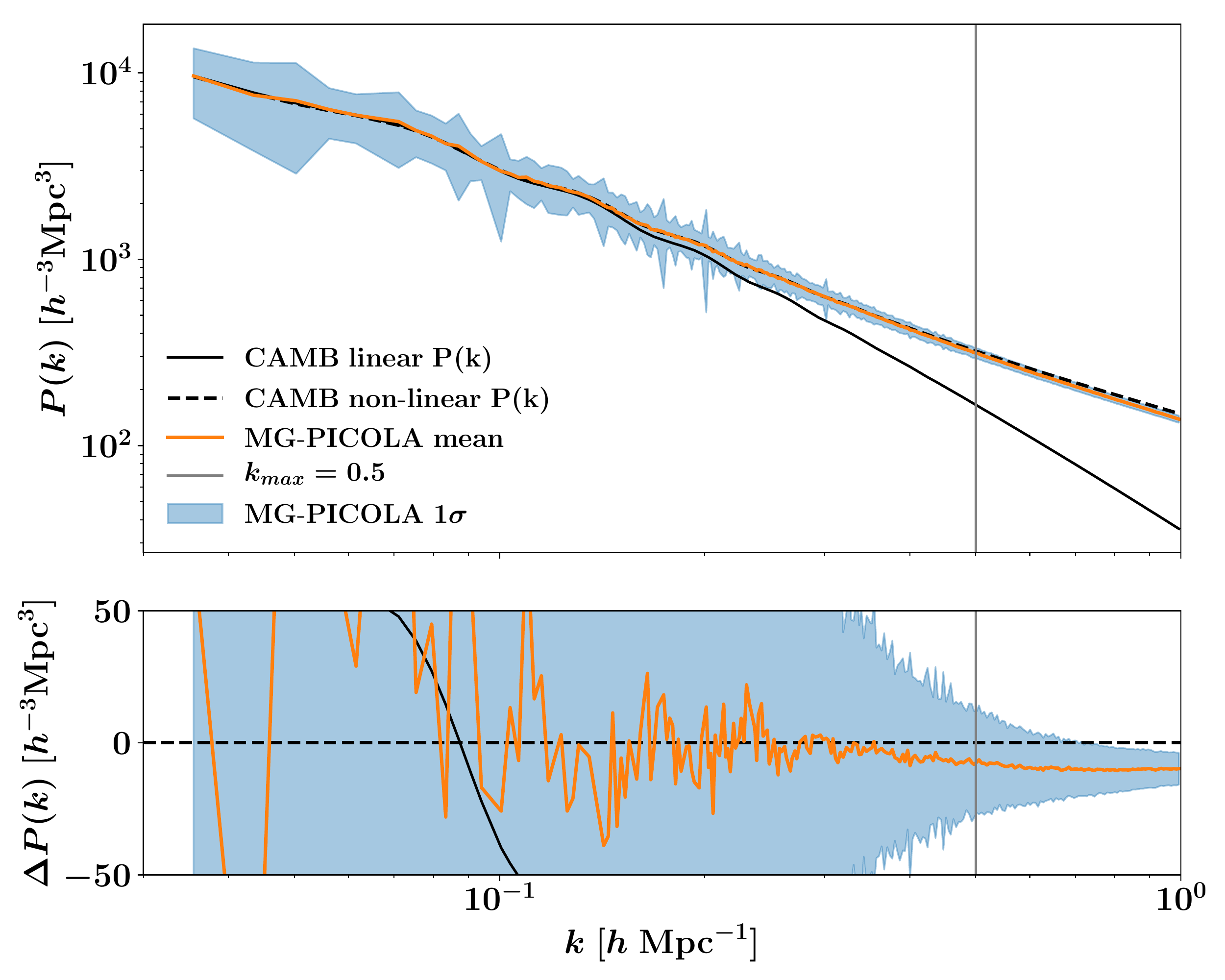}
	\caption{In the top panel we compare the mean (blue) and $1\sigma$ distributions (blue envelopes) of the power spectra calculated on dark matter particles from our fiducial suite of simulations to the linear and non-linear {\sc CAMB} power spectra. In the bottom panels we show the difference between the measured and non-linear {\sc CAMB} power spectra. The power spectra from {\sc MG-PICOLA} appears to be accurately reproduced up to about $k=0.7$, but we conservatively apply a scale cut of $k<k_{\rm max}$ where $k_{\rm max}=0.5$.}
	\label{fig:power_spec_dif_fiducial}
\end{figure}

The reliability of these simulations is evaluated by comparing the power spectrum, calculated on the dark matter particles from the fiducial suite, to the non-linear power spectrum calculated from {\sc CAMB}. We plot the 1$\sigma$ difference variation in the power spectrum in Figure \ref{fig:power_spec_dif_fiducial}. Although this test shows the simulations can be trusted up to $k < 0.7\ h {\rm Mpc}^{-1}$, we apply a conservative scale cut of $k_{\rm max} < 0.5\ h {\rm Mpc}^{-1}$ in Fourier space and $l_{\rm min}>4\upi\ h^{-1}{\rm Mpc}$ in real space.

\begin{figure*}
	\includegraphics[width=\textwidth]{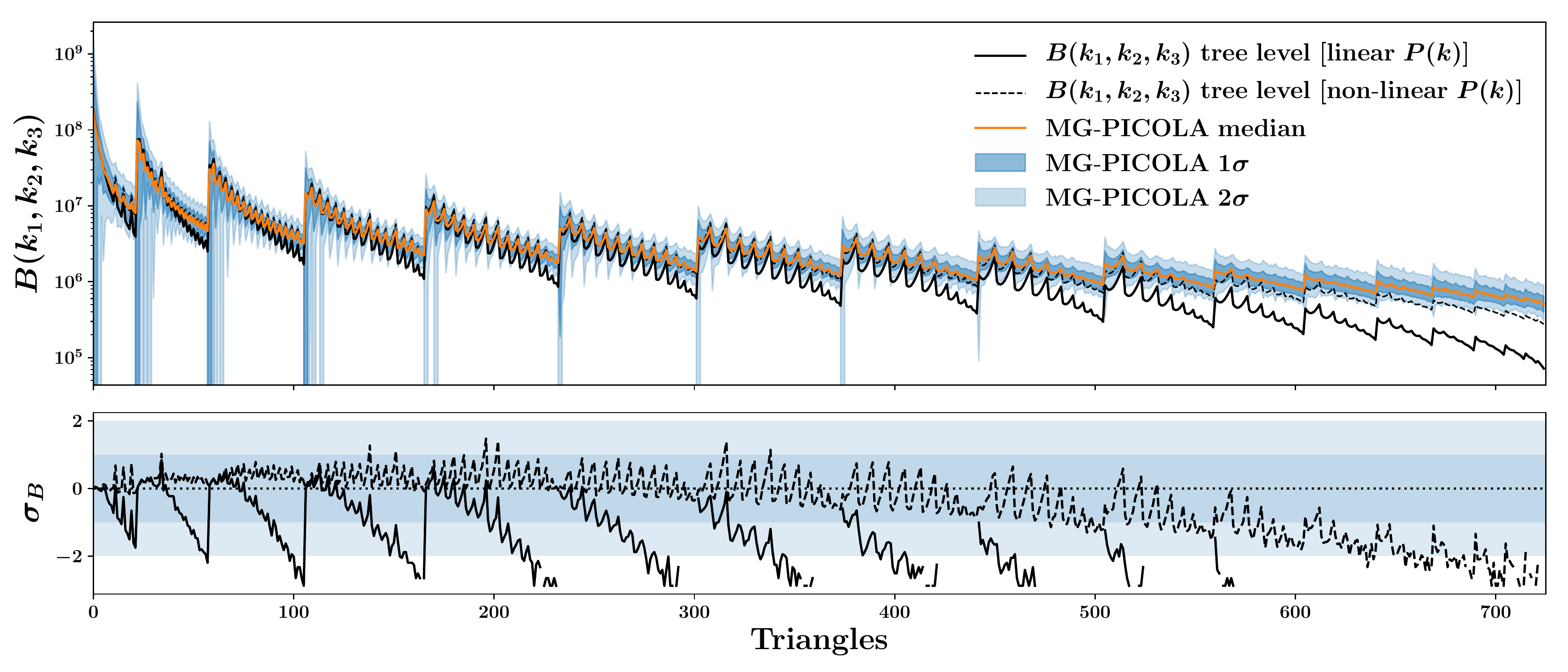}
	\caption{In the top panel we compare the mean (blue) and $1\sigma$ and $2\sigma$ distributions (blue envelopes) of the bispectrum calculated on dark matter particles (from our fiducial suite of simulations) against theoretical bispectra calculated using the linear and non-linear {\sc CAMB} power spectra . The $x$-axis displays triangle index (generated by listing triangles in lexographic order based on sides $k_{1}, k_{2}$ and $k_{3}$ where all elements are below $k_{\rm max}$). In the bottom panel we show the significance between the measured and theoretical values. The theoretical bispectrum measurements are made using \citet{DavideMethod2018} and will only be accurate up to the quasi-linear regime; since we are pushing to more non-linear scales the discreprency for smaller triangles is expected. Using the non-linear $P(k)$ for the bispectrum is an approximation that only helps in partially reducing the discrepancy between the tree-level model and the measurements by using loop corrections for the power spectrum. A better model would be given by using one-loop corrections to the bispectrum.}
	\label{fig:bispec_dif_fiducial}
\end{figure*}

\subsection{Measurements}
\label{measurements}

We use haloes from {\sc MG-PICOLA} as a proxy for galaxies. These are found using the friends-of-friends halo finder {\sc MatchMaker}\footnote{\href{https://github.com/damonge/MatchMaker}{https://github.com/damonge/MatchMaker}} which was found to be consistent (for the heaviest haloes) to the phase space halo finder {\sc Rockstar} \citep{rockstar}. Unlike $P(k)$ and $B(k_{1}, k_{2}, k_{3})$ which are unaffected by the density of tracers, the MST will exhibit different profiles purely based on the different halo counts. Since different number of haloes are produced from simulations with different cosmologies we mitigate this issue by performing our measurements on only the heaviest \num{5000} haloes. In practice such a restriction would not be imposed on $P(k)$ or $B(k_{1}, k_{2}, k_{3})$ measurements, but here we wish to simply establish whether the MST improves on the constraints of $P(k)+B(k_{1}, k_{2}, k_{3})$. 

We will explore replicating realistic survey properties in later work but in practice if we were simulating a galaxy catalogue, we would have to use a halo occupation distribution (HOD) model where we would tune the parameters of the HOD to have the same galaxy density as the actual survey. What we do here is a simplified version of that. The simulations constructed used haloes with masses between $10^{12}$ and $10^{15}\si{\solarmass}$. The number density ($\sim 3.2\times 10^{-4}\ h^{-3}{\rm Mpc^{3}}$) is similar to the BOSS LOWZ sample between redshift $0.3-0.4$ and to the CMASS sample between redshift $0.5-0.6$ \citep[see Figure 1 of][]{Tojeiro2014}. Assuming a linear bias of $b^{2} = P_{haloes}(k)/P(k)$ we found the fiducial simulations to have a bias of $b\sim1.3$; this is more similar to the bias observed in eBOSS for emission line galaxies ($b\sim1.4$) than in BOSS for luminous red galaxies ($b\sim2$).

\subsubsection{Power spectrum and bispectrum}

Power spectrum and bispectrum measurements are performed through DFT algorithms as implemented by \textsc{fftw3}.\footnote{Fastest Fourier Transform in the West, \href{http://www.fftw.org}{http://www.fftw.org}} We use the cloud-in-cell (CIC) mass assignment scheme using $64^{3}$ cartesian grid cells to define a discrete overdensity field in configuration space, later transformed into Fourier space. The size of the simulation box is $L_{\rm box}=250\,h^{-1}{\rm Mpc}$ and therefore, the mass resolution of the discrete over-density field is $\sim 3.9\, {\rm Mpc}h^{-1}$. We compute the power spectrum between the fundamental frequency, $k_{\rm f}=2\upi/L_{\rm box}$, and a maximum frequency, $k_{\rm max}=0.5\,h{\rm Mpc}^{-1}$, in bins of $k_{\rm f}$.

The power spectrum and bispectrum measurements are performed using the code and estimator described in \cite{Hector2017}. For the bispectrum we initially perform the measurements in bin sizes of $k_{f}$. In this case we ensure that the three $k$-vectors of the bispectrum form closed triangles, and without loss of generality we define $k_1\leq k_2 \leq k_3$. We include all the closed triangles with $k_3<k_{\rm max}$. The bispectrum data vector, $B(k_1,k_2,k_3)$, contains around 700 elements. In Figure \ref{fig:bispec_dif_fiducial}, the bispectra measured on dark matter particles from the fiducial simulations are compared to theoretical values, showing good agreement until we reach non-linear regimes where the theory can no longer be trusted.

Using measurements of the power spectrum and bispectrum on the haloes of the fiducial suite, we were able to determine the skewness and kurtosis of the individual elements of the data vector. We found that elements with $k < 0.125\ h{\rm Mpc}^{-1}$ contained much higher than expected skewness and kurtosis (i.e. exceeded the expected skewness and excess kurtosis of a Gaussian data set by $2\sigma$) and as such we limit the power spectrum and bispectrum measurements to $k > 0.125\ h{\rm Mpc}^{-1}$. This reduced the bispectrum data vector from $\sim 700$ to $\sim 500$. We then use a maximal compression technique (based on the work of \citealt{Tegmark1997} and \citealt{Moped}) to compress the bispectrum data vector to three elements \citep[following][]{DavideMethod2018, davideboss2018}. Such a compression allows us to estimate the covariance matrix for a number of triangle configurations much larger than the number of available simulations.

\subsubsection{Minimum spanning tree}

The MST measurements are made with a scale cut of $l_{\rm min} > 4\upi\ h^{-1}{\rm Mpc}$, which corresponds to the wavelength ($\lambda=2\upi/k$) of the largest $k$ modes ($k_{\rm max}$) probed by $P(k)$ and $B(k_{1}, k_{2}, k_{3})$. The MST statistics are then binned, which presents a problems as counts are discrete. For large counts, the distribution can be approximated by a Gaussian and as such we only select bins which we found the mean of our fiducial data vectors to have counts of greater than $50$.

\begin{figure*}
	\begin{minipage}[b]{0.49\textwidth}
		\includegraphics[width=\textwidth]{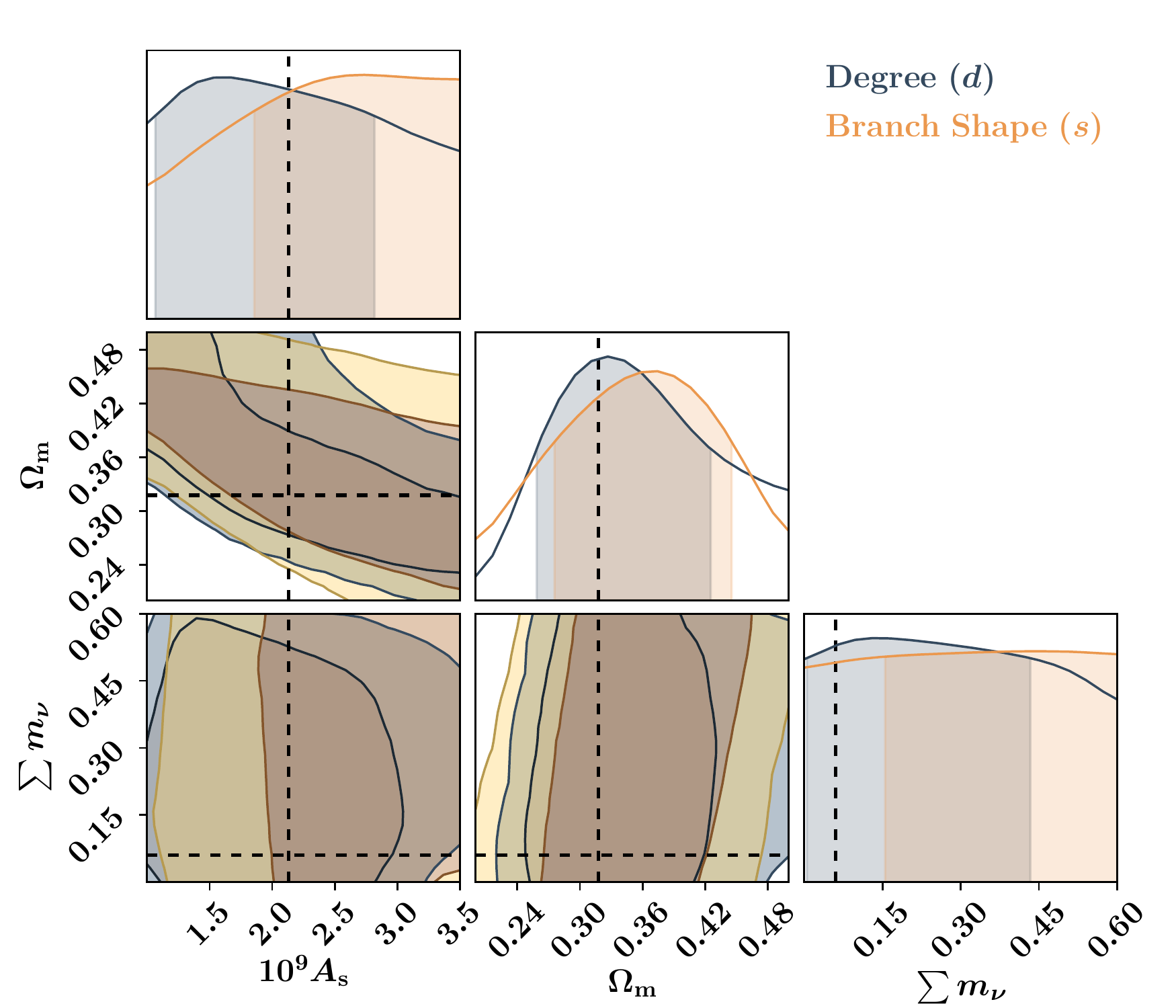}
	\end{minipage}
	\begin{minipage}[b]{0.49\textwidth}
		\includegraphics[width=\textwidth]{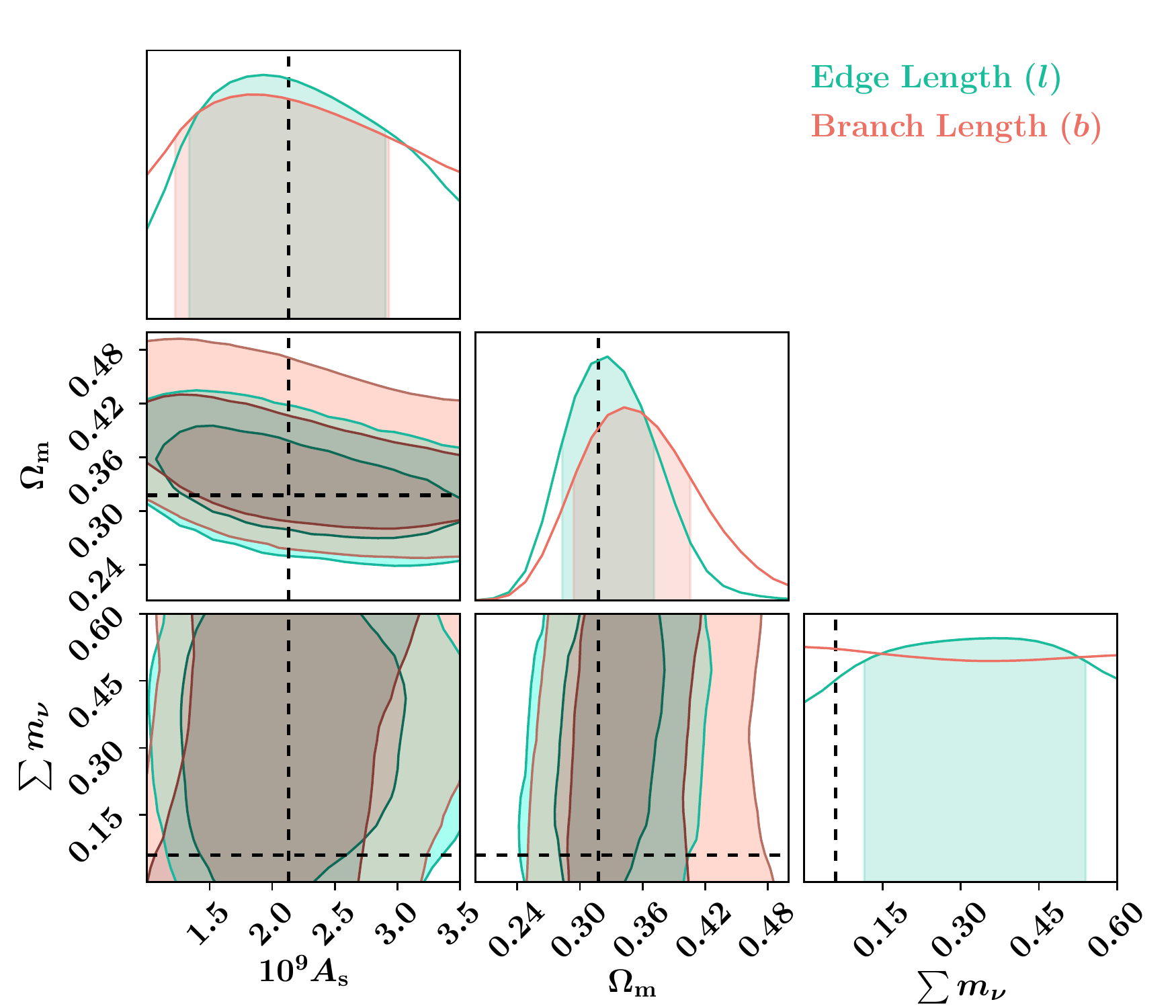}
	\end{minipage}
	\caption{Posterior distributions on cosmological parameters as constrained by the individual components of the MST. On the left, we show those from the degree and branch shape and on the right from edge and branch lengths. Branch shapes are the least sensitive, whilst the degree gives broad constraints but rules out parts of the parameter space. Edge and branch length show similar posterior distributions with tighter constraints coming from edges.}
	\label{fig:mst_components_dlbs}
\end{figure*}

\subsection{Parameter estimation}
\label{param_estimation}

Using the noisy estimates of the theory $\mathbi{d}_{\rm Grid}$ (the mean of five grid simulations at each point in parameter space) we can interpolate using GPs (see Appendix \ref{GPinterp}) from a $6\times 6\times 6$ to a $20\times 20\times 20$ grid with theoretical data vectors $\pmb{\mu}_{GP}$ and uncertainty $\pmb{\sigma}_{GP}$ which is used instead of an MCMC due to the low dimensionality of the parameters.  The sample covariance matrix, $\mathbfss{S}$, is estimated from 400 fiducial simulations (the other 100 fiducial simulations are used to apply a coverage correction, \citealt{Sellentin2019}). The posterior for each of our ten mocks, denoted by the data vector $\mathbi{d}$, is evaluated using the likelihood function (which accounts for an estimated sample covariance, see \citealt{SellentinHeavens2016}; \citealt{Jeffrey2018})
\begin{equation}
	\mathcal{L}(\mathbi{d} | \pmb{\theta}) \propto {\rm det}(\mathbfss{C})^{-1/2}\left[1+\frac{(\mathbi{d}-\pmb{\mu}_{\rm GP})^{\top}\cdot\mathbfss{C}^{-1}\cdot(\mathbi{d}-\pmb{\mu}_{\rm GP})}{N-1}\right]^{-\frac{N}{2}},
\end{equation}
where the uncertainty in the GPs regression is added to the sample covariance, i.e. $\mathbfss{C}=\mathbfss{S}+ \mathbfss{S}_{\rm GP}$, where elements of $(\mathbfss{S}_{\rm GP})_{\rm ij} = \pmb{\sigma}_{\rm GP, i}\pmb{\sigma}_{\rm GP, j}\delta_{k}(v_{\rm i}, v_{\rm j})$ where $\delta_{k}$ is the Kronecker delta function and $v_{\rm i}$ and $v_{\rm j}$ are only equal if the same GPs hyperparameters were used to construct these elements of the data vector \citep[following][which assume maximal dependency between elements of the data vector constructed from the same GPs hyperparameters]{Bird2019,Keir2018}.

Finally we apply a coverage correction \citep{Sellentin2019} using 100 fiducial simulations not included in the calculation of the covariance matrix. This accounts for unrecognized sources of biases. We found that all methods exhibited overconfident confidence contours. For $P(k)$ and $B(k_{1}, k_{2}, k_{3})$ this is believed to have arisen due to non-Gaussian features in the data set. Although we attempted to limit this by selecting regions of the data vector that had fairly low skewness and kurtosis, we found that the skewness for $P(k)$ tended to be consistently positive, whilst the excess kurtosis for the maximally compressed $B(k_{1}, k_{2}, k_{3})$ was always $> 1\sigma$ than expected if the data were Gaussian. For the MST, this effect is larger which we suspect occurs due to two reasons: (1) similar to $P(k)$ and $B(k_{1}, k_{2}, k_{3})$ the data vector is non-Gaussian and (2) the scale cut adds an extra stochasticity to the data vector that is not fully captured by the covariance matrix.

\begin{figure}
	\includegraphics[width=\columnwidth]{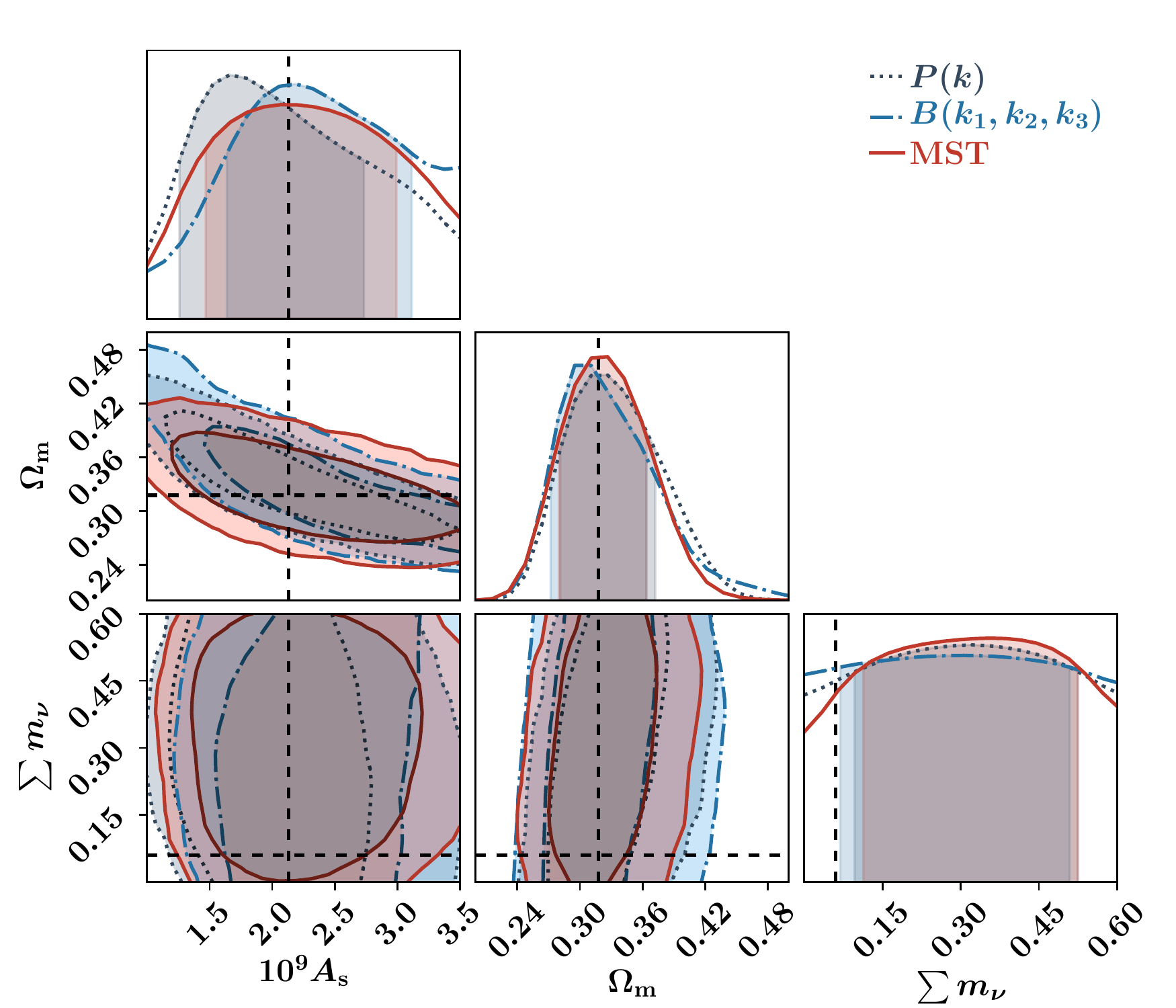}
	\caption{The posterior distributions are shown for power spectrum ($P(k)$, shown in grey), bispectrum ($B(k_{1}, k_{2}, k_{3})$, shown in blue) and MST (shown in red). The tightest constraints on $A_{s}$ and $\Omega_{m}$ are given by the MST whilst $B(k_{1}, k_{2}, k_{3})$ provides better constraints on $\sum m_{\nu}$.}
	\label{fig:pkbkmst}
\end{figure}

\begin{figure}
	\includegraphics[width=\columnwidth]{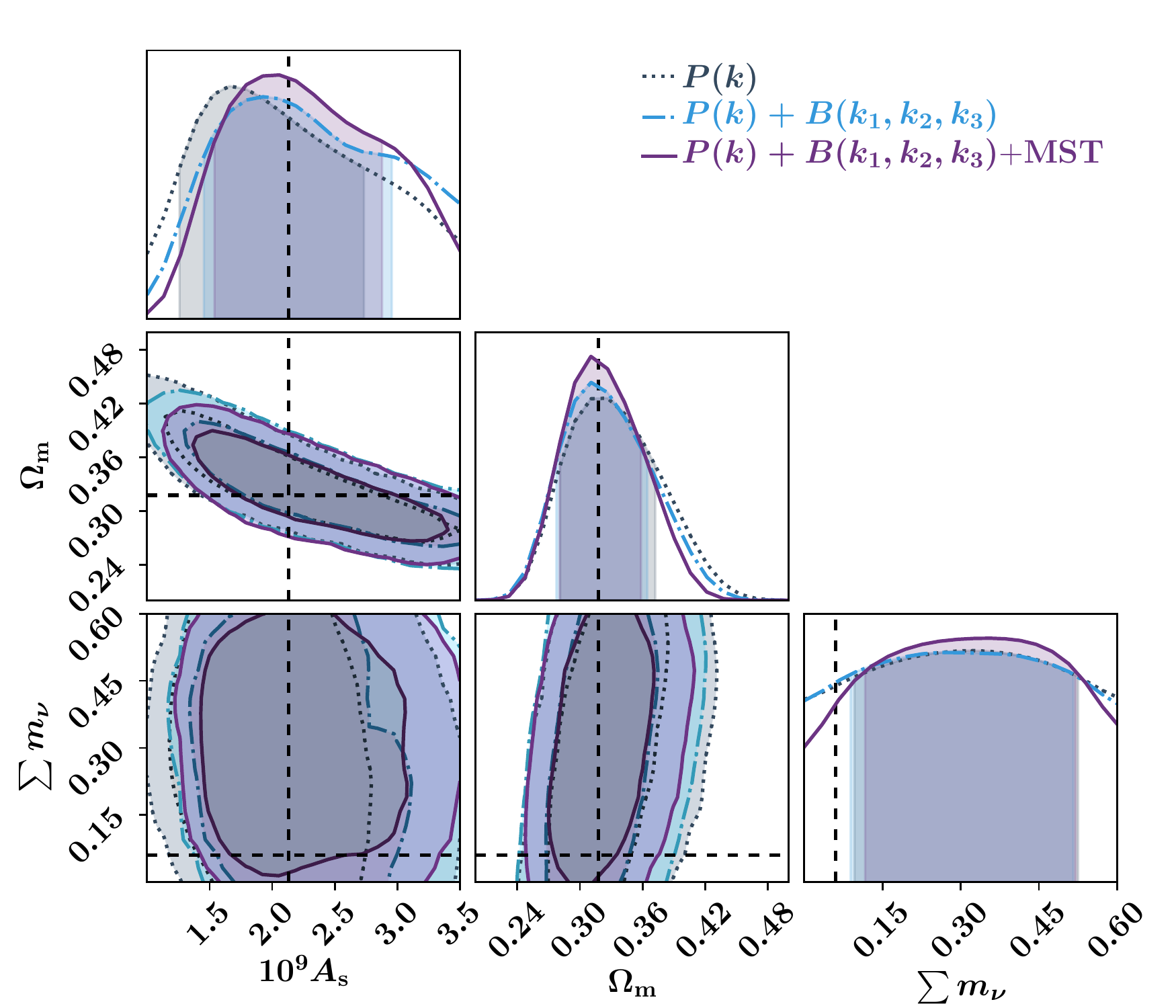}
	\caption{The posterior distributions for cosmological parameters as constrained by (a) power spectrum ($P(k)$, shown in dark grey) (b) power spectrum and bispectrum ($P(k) + B(k_{1}, k_{2}, k_{3})$, shown in blue) and (c) power spectrum, bispectrum, and MST ($P(k)+B(k_{1}, k_{2}, k_{3})+{\rm MST}$, shown in purple). }
	\label{fig:pkbkmst_combos}
\end{figure}

\begin{figure}
	\includegraphics[width=\columnwidth]{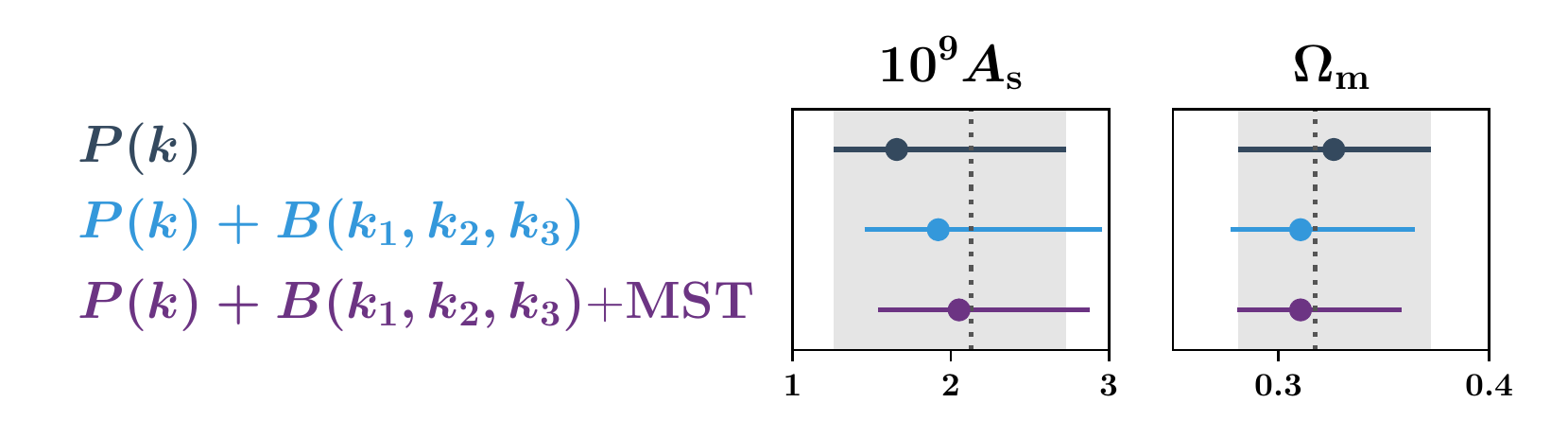}
	\caption{The $1\sigma$ constraints on $A_{\rm s}$ and $\Omega_{\rm m}$ are shown for $P(k)$ (dark gray), $P(k)+B(k_{1}, k_{2}, k_{3})$ (blue) and $P(k)+B(k_{1}, k_{2}, k_{3})+{\rm MST}$ (purple). This plot shows how including the MST improves constraints on $A_{\rm s}$ by $\sim 12\%$ ($\sim 10\%$) and on $\Omega_{\rm m}$ by $\sim 17\%$ ($\sim 12\%$) with respect to $P(k)$ ($P(k)+B(k_{1}, k_{2}, k_{3})$.}
	\label{fig:pkbkmst_summary}
\end{figure}

\subsection{Comparison}

The posterior distributions are measured for the three statistics and their combinations. Correlations between each statistic are accounted for by using a covariance matrix that is not block diagonal. In Figures \ref{fig:mst_components_dlbs}, \ref{fig:pkbkmst} and \ref{fig:pkbkmst_combos} we show the posterior distributions measured on the mean of the data vectors from 10 mocks allowing for better visual comparison of the errors whilst improvement in parameter constraints are stated according to the average improvement when measured on the mocks independently.

\subsubsection{Components of the minimum spanning tree}

We compare the constraints from the four individual components of the MST. The elements of the MST statistics are counts, and as such they follow a Poisson distribution. We apply a cut on the data vector based on where the mean of the fiducial MST statistics had counts $>50$, where expect the Poisson distribution to be approximately characterized by a Gaussian. In Figure \ref{fig:mst_components_dlbs}, we display the constraints from the individual components of the MST. Of the four statistics $s$ is the least constraining and provides very little information; this is followed by $d$ which, although it has very broad posteriors, appears at least to rule out parts of the parameter space (low $A_{\rm s}$, $\Omega_{\rm m}$ and high $\sum m_{\nu}$). The MST statistics $l$ and $b$ provide constraints having similar degeneracies with $l$ providing somewhat tighter constraints.

\subsubsection{$P(k)$, $B(k_{1}, k_{2}, k_{3})$, and MST}

In Figure \ref{fig:pkbkmst} we compare the constraints from $P(k)$, $B(k_{1}, k_{2}, k_{3})$ and MST. All three appear to have similar degeneracies and as such are unable to establish meaningful constraints on $A_{\rm s}$ and $\sum m_{\nu}$. The constraints on $\Omega_{\rm m}$ are more conclusive but are fairly similar. The constraints on $\sum m_{\nu}$ tend to show a broad peak towards the centre of the prior range. Since the constraints on neutrino mass are poor the kernel-length scale for $\sum m_{\nu}$ of the GPs is quite broad and as such the estimates of the theory vector are smoother in the centre. This creates a slight bias towards the centre of the parameter space. This effect is also seen in Figure \ref{fig:pkbkmst_combos}.

\subsubsection{Combining $P(k)$, $B(k_{1}, k_{2}, k_{3})$ and MST}

In Figure \ref{fig:pkbkmst_combos}, we combine the statistics and compare their relative constraints which is more clearly shown in Figure \ref{fig:pkbkmst_summary}. In combining $P(k)$ and $B(k_{1}, k_{2}, k_{3})$, we find an improvement of $\sim 6\%$ in the constraints of $\Omega_{\rm m}$ and $\sim 3\%$ for $A_{\rm s}$. When combined with the MST the constraints on $\Omega_{\rm m}$ improve by $\sim 17\%$ and on $A_{\rm s}$ improve by $\sim 12\%$ with respect to $P(k)$ ($\sim 12\%$ for $\Omega_{\rm m}$ and $\sim 10\%$ for $A_{\rm s}$ with respect to $P(k)+B(k_{1}, k_{2}, k_{3})$). Since we have ensured the same scale cuts, i.e. $k_{\rm max} = 0.5\ h{\rm Mpc}^{-1}$ for $P(k)$ and $B(k_{1}, k_{2}, k_{3})$ and $l_{\rm min}=4\upi\ h^{-1}{\rm Mpc}$, we can be fairly certain that the additional information is not coming from the MST having access to smaller scales. Furthermore, the maximally compressed $B(k_{1}, k_{2}, k_{3})$ has been shown by \citet{DavideMethod2018} to improve parameter constraints by allowing the inclusion of many more triangle configurations than standard bispectrum analysis. Therefore, we can be fairly certain that the additional information is coming from the MST's detection of patterns in the cosmic web, information which would be present in higher order functions such as the trispectrum, thus confirming the heuristic arguments made in Section \ref{heuristic}.

%% file: discussion.tex
\section{Discussion}
\label{discussion}

In this paper, we have sought to understand whether the MST can be used for parameter inference in cosmology. Until now, the MST has been predominantly used to search for large-scale features. This type of information has largely been overlooked as traditionally two-point statistics are completely insensitive to phase information. In constructing the MST we hope to pick up patterns in the cosmic web and use this to improve parameter constraints.

In Section \ref{higherorder}, we argue heuristically why the MST should be sensitive to higher order statistics (i.e. three-point and beyond). This is demonstrated using simulated galaxies (from the Illustris $N$-body simulation) and a random walk simulation (produced using an adjusted L\'{e}vy Flight algorithm) with virtually identical 2PCF by design but different higher order statistics.

In Section \ref{systematics}, we look at the effects of boundaries and masks, RSD and scale cuts. Boundaries and masks\footnote{Boundaries can be thought of as a survey's footprint, whilst the mask would also include holes and varying completeness levels.} tended to produce longer edge lengths, whilst the degree and branch shape appeared to be unaffected. RSD is shown to have a significant impact on the MST statistics and thus should be incorporated in any future study. Lastly, we develop a strategy to impose a scale cut on the MST. This is done by removing edges below a set length in the $k$NN graph and then constructing the MST from this. Unfortunately this creates some artefacts in the degree and branch shape distributions. It is also believed that this method distorts some of the information we are trying to learn. As such alternatives or improvements to this method should be explored.

In Section \ref{section_nbody}, we look to determine what the MST actually measures, finding the MST to be highly sensitive to its local density. This is demonstrated by the fact that nodes in overdensities tended to have a degree of 2.

Lastly in Section \ref{comparison}, we determine whether the MST provides information not present in power spectrum and bispectrum. We do this by obtaining parameter constraints on $A_{\rm s}$, $\Omega_{\rm m}$, and $\sum m_{\nu}$ for 10 halo mock catalogues. To keep the density of haloes the same in all our simulations we use only the most massive 5000 haloes and measure the power spectrum $P(k)$, bispectrum $B(k_{1}, k_{2}, k_{3})$ and MST statistics. The individual methods provided similar constraints although due to the degeneracies with $\Omega_{\rm m}$ we were unable to obtain meaningful constraints on $\sum m_{\nu}$. We found that combining the three methods narrows the $1\sigma$ constraints on $\Omega_{\rm m}$ by $\sim 17\%$ and on $A_{\rm s}$ by $\sim 12\%$ with respect to $P(k)$ and $\sim 12\%$ on $\Omega_{\rm m}$ and $\sim 10\%$ on $A_{\rm s}$ with respect to $P(k) + B(k_{1}, k_{2}, k_{3})$, thus showing that the MST is providing information not present in the power spectrum or bispectrum. We expect this to improve with improved implementation of scale cuts and greater statistical power from larger samples.

The MST provides several advantages over existing methods but has some important limitations. The main advantages are: (1) it is sensitive to patterns in the cosmic web and (2) the algorithm is computationally inexpensive. The naive brute force implementation of $N$-point statistics for $n$ points is an $\mathcal{O}(n^N)$ process. While there exist faster implementations of the 2PCF and 3PCF \cite[see][]{Scoccimarro2015, Zachary2016} there are no such methods for higher order statistics. On the other hand, the MST is sensitive to higher order statistics and the Kruskal algorithm used here is approximately an $\mathcal{O}(n\log n)$ process. In the MST, we have a window into these higher order statistics but at a fraction of the computational cost. The main limitations of the MST: (1) we need simulations to estimate the statistics and (2) the statistic is dependent on the density of the tracer. This means we will need to create simulations that both match the survey properties as well as the density of the tracers used.

In future work we look to apply the MST to current and future galaxy redshift surveys. In doing so we hope to better understand how to implement scale cuts and mitigate any of the resulting effects that occur as a result. One thing we have not studied in this paper is the effect of galaxy bias which should be explored in future. This could be achieved by varying HOD parameters. Lastly, ML algorithms and AI are powerful new tools to cosmology \citep[see][]{ho2017,Tomasz2018}, however it is difficult to gain an intuition into what these algorithms are learning. Since the MST is relatively simple this could be used to gain insight into this work, providing a bridge between the traditional two-point and a full ML/AI approach.

Finally, the MST statistics presented in this paper have been produced by the {\sc Python} module {\sc MiSTree} \citep{mistree}, which implements the procedures detailed in Section \ref{method}. The module is made publicly available (see \href{https://github.com/knaidoo29/mistree}{https://github.com/knaidoo29/mistree} for documentation) and can handle data sets provided in 2D and 3D Cartesian coordinates, spherical polar coordinates and coordinates on a sphere (either celestial RA, Dec. or simply longitude and latitude).

%% file: appendix.tex
\section{Gaussian process interpolation}
\label{GPinterp}

We will be modelling data vectors following a method similar to that of \citet{Keir2018} and \citet{Bird2019} in which they emulated the 1D flux power spectrum of the Lyman-$\alpha$ forest using GPs. In this section we provide a brief introduction to GPs and outline their usage in this paper. A comprehensive overview of GPs and their applications can be found in \citet{gp2006}, while an overview of their implementations for vectors can be found in \citet{vectorgp2011}. 

\subsection{Introduction}

GPs are a non-parametric kernel-based regression and interpolation method. In GPs we model the desired function $f(x)$ as a stochastic process with a prior probability over all parametric functions. For a finite input data set $\pmb{X} =\{x_{\rm 1}, ..., x_{\rm n}\}$, this can be modelled as a multivariate Gaussian,
\begin{equation}
\label{gp_gaussian_eq}
\mathcal{GP} = \mathcal{N}\left(\pmb{m}(\pmb{X}), \mathbfss{K}\left(\pmb{X},\pmb{X}'\right)\right),
\end{equation}
with mean $\pmb{m}(\pmb{X})$ and covariance $\mathbfss{K}(\pmb{X},\pmb{X}')$. Given training data $\pmb{Y}_{1}$ at $\pmb{X}_{1}$, we model the posterior of the function $f(x)$ at new positions $\pmb{X}_{2}$ as a multivariate Gaussian,
\begin{equation}
P\left(\pmb{Y}_{2} | \pmb{X}_{1}, \pmb{Y}_{1}, \pmb{X}_{2}\right) = \mathcal{N}\left(\pmb{\mu}_{2|1}, \mathbfss{S}_{2|1}\right),
\end{equation}
with mean $\pmb{\mu}_{2|1}$ and covariance $\mathbfss{S}_{2|1}$. Assuming that both $\pmb{Y}_{1}$ and $\pmb{Y}_{2}$ are drawn from the same multivariate Gaussian, as our prior on the function indicates (see Equation \ref{gp_gaussian_eq}), we can write the relation
\begin{equation}
	\begin{bmatrix} \pmb{Y}_{1}\\ \pmb{Y}_{2} \end{bmatrix} \sim \mathcal{N}\left(\begin{bmatrix} \pmb{\mu}_{1}\\ \pmb{\mu}_{2} \end{bmatrix},\begin{bmatrix}\mathbfss{K}_{11} + \mathbfss{I}\sigma_{\rm n}^{2} & \mathbfss{K}_{12} \\ \mathbfss{K}_{21} & \mathbfss{K}_{22}\end{bmatrix}\right),
\end{equation}
where $\mathbfss{I}$ is the identity matrix and $\sigma_{\rm n}$ is the standard deviation of the training data $\pmb{Y}_{1}$ (which is either known or fitted later). Thus assuming the mean function is zero we arrive at the predicted mean and covariance,
\begin{equation}
\pmb{\mu}_{2|1}=\left[\left(\mathbfss{K}_{11}  + \mathbfss{I}\sigma_{\rm n}^{2}\right)^{-1}\mathbfss{K}_{12}\right]^{\top}\pmb{Y}_{1},
\end{equation}
\begin{equation}
	\mathbfss{S}_{2|1} = \mathbfss{K}_{22} - \left[\left(\mathbfss{K}_{11} + \mathbfss{I}\sigma_{\rm n}^{2}\right)^{-1}\mathbfss{K}_{12}\right]^{\top}\mathbfss{K}_{12},
\end{equation}
where the dependence on $\mathbfss{K}_{21}$ has been removed due to the symmetry $\mathbfss{K}_{12}=\mathbfss{K}_{21}^{\top}$. Note that in practice we determine the GPs mean and standard deviation at a single new position and thus the standard deviation is simply a scalar -- this means that $\mathbfss{K}_{12}$ and $\mathbfss{K}_{21}$ reduce to vectors and $\mathbfss{K}_{22}$ to a scalar.
 
\subsection{Kernel}

GPs use kernels to weight the interdependency of points in parameter space. In our model we use a Gaussian kernel,
\begin{equation}
	\kappa(\theta_{\rm i}, \theta_{\rm j}) = \sigma_{{\rm GP}}^{2} \exp \left(-\frac{r^{2}}{2}\right).
\end{equation}
Here,
\begin{equation}
	r = \frac{|\theta_{\rm i,1}-\theta_{\rm j,1}|^{2}}{2l_{{\rm GP,1}}^{2}}+\frac{|\theta_{\rm i,2}-\theta_{\rm j,2}|^{2}}{2l_{{\rm GP,2}}^{2}}+\frac{|\theta_{\rm i,3}-\theta_{\rm j,3}|^{2}}{2l_{{\rm GP,3}}^{2}};
\end{equation}
$\sigma_{{\rm GP}}$, $l_{{\rm GP, 1}}$, $l_{{\rm GP, 2}}$, and $l_{{\rm GP, 3}}$ are GPs hyperparameters to be fitted with independent scale terms for each axis in the parameter space; and $\pmb{\theta} = \left[10^{9}A_{\rm s}, \Omega_{\rm m}, m_{\nu}\right]$.  The covariance matrix $\mathbfss{K}$ is then defined to have elements
\begin{equation}
(\mathbfss{K})_{\rm ij}= \kappa (\theta_{\rm i}, \theta_{\rm j}) + \sigma_{\rm n}^{2}\delta_{k}(\theta_{\rm i}, \theta_{\rm j}),
\end{equation}
with an additional noise term $\sigma_{\rm n}$.

\subsection{Hyperparameter optimization}

The hyperparameters $\pmb{\phi} = [\sigma_{\rm GP}, l_{\rm GP,1}, l_{\rm GP,2}, l_{\rm GP,3}]$ are optimized by maximising the likelihood function
\begin{equation}
	\mathcal{L}(\pmb{D} | \pmb{\theta}, \pmb{\phi}) = \sum_{i}^{n}\mathcal{L}(\pmb{d}_{\rm i}| \pmb{\theta},\pmb{\phi}),
\end{equation}
where $\pmb{D}$ are the ensemble of training data vectors, $\pmb{d}_{\rm i}$ is an element of a specific data vector and
\begin{equation}
	\mathcal{L}(\pmb{d}_{\rm i}|\pmb{\theta}, \pmb{\phi})=-\frac{1}{2}\pmb{d}_{\rm i}^{\top}\mathbfss{K}^{-1}\pmb{d}_{\rm i} - \frac{1}{2}\log|\mathbfss{K}|-\frac{n}{2}\log 2\upi.
\end{equation}

\subsection{Implementation and validation}

\begin{figure}
	\centering
	\includegraphics[width=0.9\columnwidth]{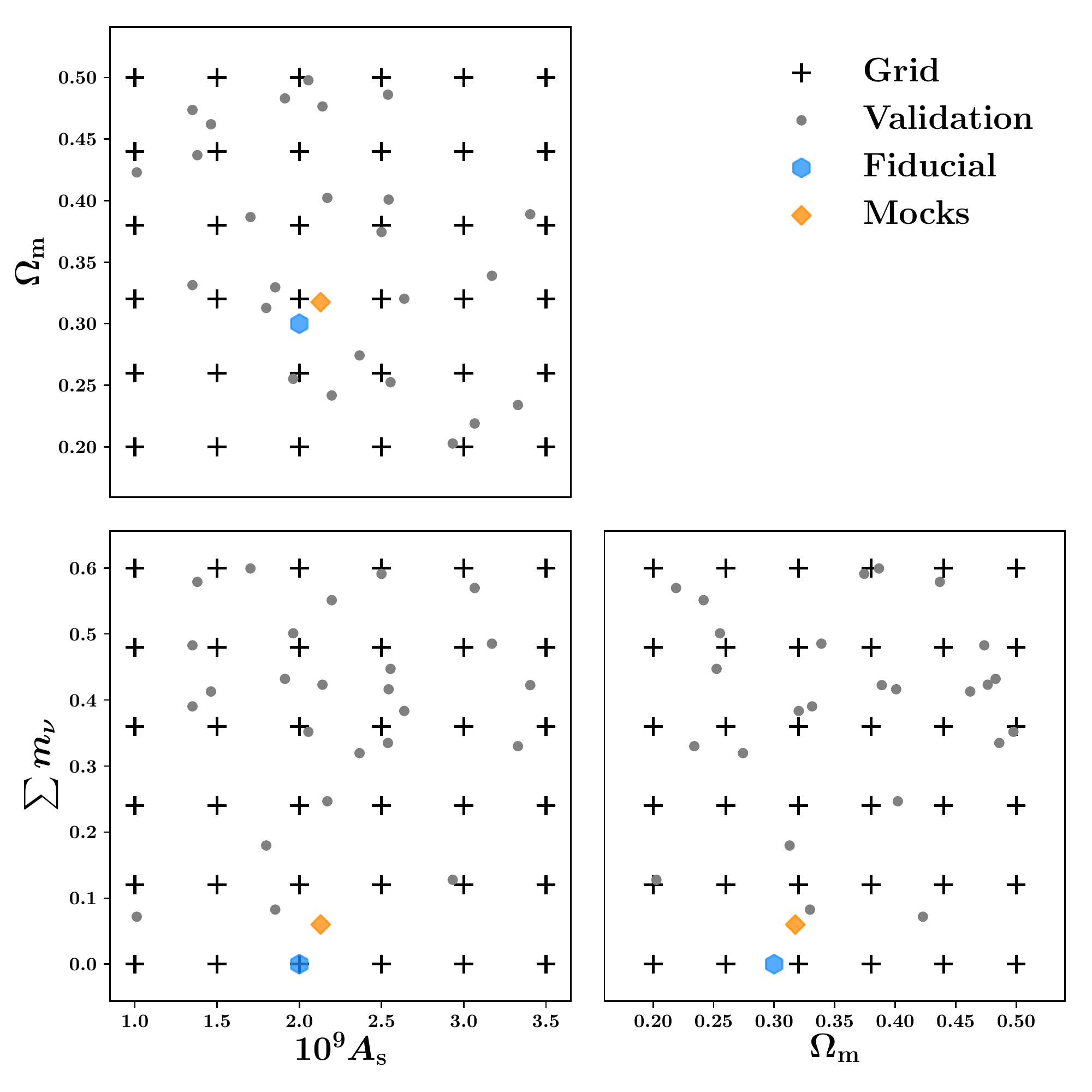}
	\caption{The positions in parameter space of simulations (grid, validation, fiducial and mocks) used in Section \ref{comparison}. Note that for the Grid simulations each cross marks the point of five simulations.}
	\label{fig:simulations}
\end{figure}

The GPs hyperparameters are trained on the measurements of $P(k)$, the maximally compressed $B(k_{1}, k_{2}, k_{3})$ and the MST statistics $d$, $l$, $b$, and $s$  (see Section \ref{measurements} for further details on these measurements) from the Grid simulations separately. In Figure \ref{fig:simulations} we show the placement of the grid, fiducial, mock and validation (used only in this section) simulations in parameter space.
\begin{figure}
	\centering
	\includegraphics[width=0.9\columnwidth]{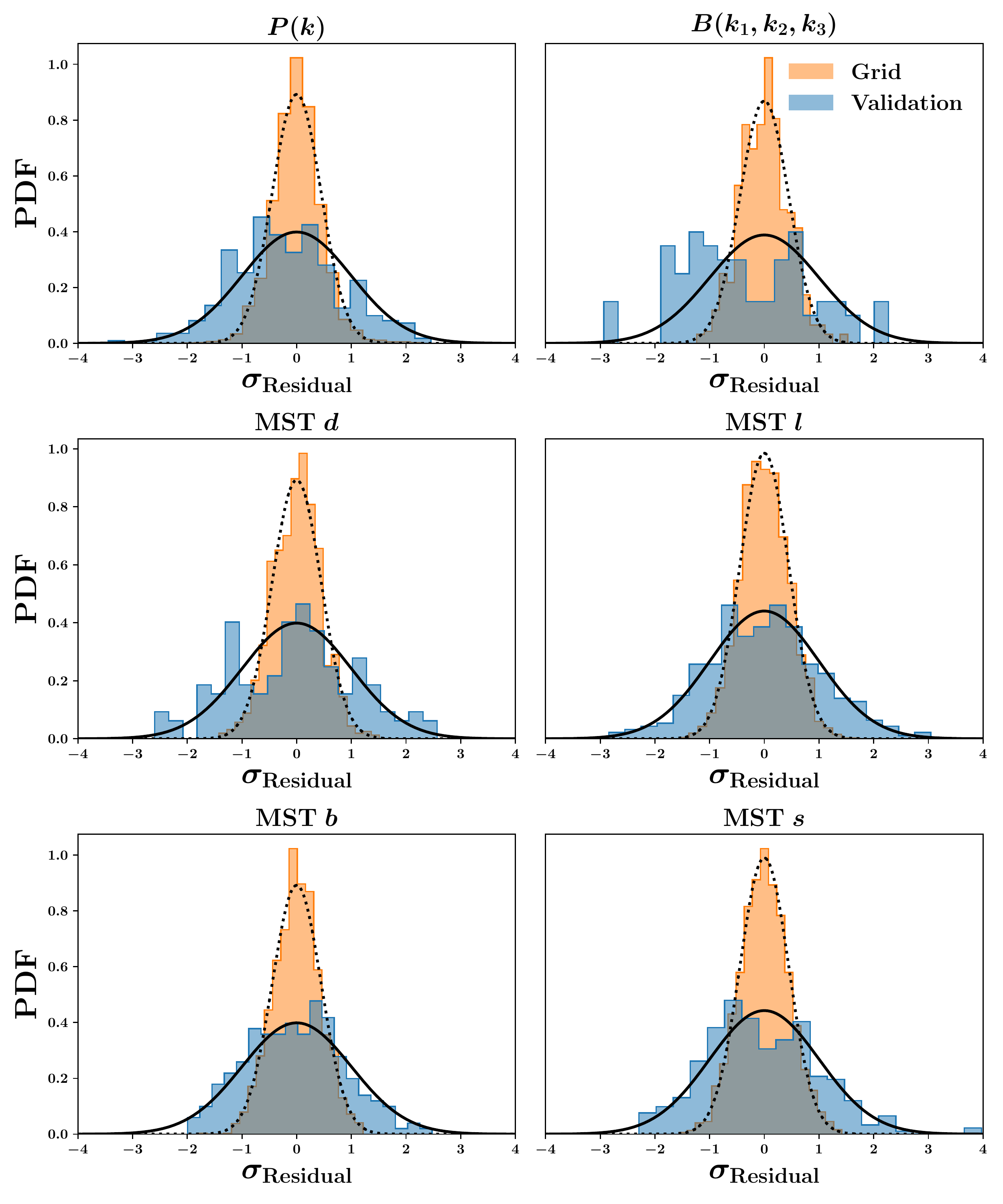}
	\caption{The residuals between the statistics of $P(k)$ (top left), maximally compressed $B(k_{1}, k_{2}, k_{3})$ (top right), MST degree (middle left), edge length (middle right), branch length (bottom left) and branch shape (bottom right) for the grid (shown by the orange histograms) and validation (shown by the blue histograms) simulations calculated from Equation \ref{residual}. Since the grid data vectors are the mean of five realizations the residuals are expected to follow a normal distribution of $\mathcal{N}(0, 1/\sqrt{5})$ (shown by the dotted black line), whilst the validation data vector are expected to follow a normal distribution of $\mathcal{N}(0, 1)$. We see that for most of the statistics the agreement is fairly good, with the exception of $B(k_{1}, k_{2}, k_{3})$ which shows more spread than is expected.}
	\label{fig:validation}
\end{figure}
To test that our GPs interpolation is emulating the statistics accurately we calculate the residuals between the grid simulations (using the mean of five realisations made at each point in parameter space),
\begin{equation}
\label{residual}
	\pmb{\sigma}_{\rm Residual} = \frac{\mathbi{d} - \pmb{\mu}_{\rm GP}}{\sqrt{\pmb{\sigma}_{\rm Fiducial}^{2} + \pmb{\sigma}_{\rm GP}^{2}}},
\end{equation}
where $\pmb{\mu}_{\rm GP}$ and $\pmb{\sigma}_{\rm GP}$ are the GPs mean and standard deviation evaluated at the same points in parameter space as $\mathbi{d}$. We plot histograms of the residuals for the grid data vectors in Figure \ref{fig:validation} shown in orange. Notice that since the grid simulations are the mean of five simulations the distribution follows a Gaussian with mean $0$ and standard deviation $1/\sqrt{5}$ (illustrated by the black dotted line). Furthermore to test that our GPs interpolation produces a good fit to simulations not present in the training data, we generate 25 new simulations (called the validation simulations) with randomly drawn cosmological parameters (shown in Figure \ref{fig:simulations}). We then again compare the residuals to that of our GPs interpolation and find a good agreement (with the exception of $B(k_{1}, k_{2}, k_{3})$) with a Gaussian with mean $0$ and standard deviation $1$ illustrated by the black full lines. 